\documentclass[twocolumn]{aastex631}
\usepackage{booktabs}
\usepackage{amssymb}
\usepackage{graphicx}

\shorttitle{Magellan/IMACS spectroscopy of Eri IV and Cen I}
\shortauthors{Heiger et al.}

\graphicspath{{./}{}}

\begin{document}

\title{Reading Between the (Spectral) Lines: Magellan/IMACS spectroscopy of the Ultra-faint Dwarf Galaxies Eridanus IV and Centaurus I}

\author[0000-0002-2446-8332]{M.~E.~Heiger}
\affiliation{Department of Astronomy and Astrophysics, University of Toronto, 50 St. George Street, Toronto ON, M5S 3H4, Canada}
\affiliation{Dunlap Institute for Astronomy \& Astrophysics, University of Toronto, 50 St George Street, Toronto, ON M5S 3H4, Canada}
\author[0000-0002-9110-6163]{T.~S.~Li}
\affiliation{Department of Astronomy and Astrophysics, University of Toronto, 50 St. George Street, Toronto ON, M5S 3H4, Canada}
\affiliation{Dunlap Institute for Astronomy \& Astrophysics, University of Toronto, 50 St George Street, Toronto, ON M5S 3H4, Canada}
\author[0000-0002-6021-8760]{A.~B.~Pace}
\affiliation{McWilliams Center for Cosmology, Carnegie Mellon University, 5000 Forbes Ave, Pittsburgh, PA 15213, USA}
\author[0000-0002-4733-4994]{J.~D.~Simon}
\affiliation{Observatories of the Carnegie Institution for Science, 813 Santa Barbara St., Pasadena, CA 91101, USA}
\author[0000-0002-4863-8842]{A.~P.~Ji}
\affiliation{Department of Astronomy \& Astrophysics, University of Chicago, 5640 S Ellis Avenue, Chicago, IL 60637, USA}
\affiliation{Kavli Institute for Cosmological Physics, University of Chicago, Chicago, IL 60637, USA}
\author[0000-0002-7155-679X]{A.~Chiti}
\affiliation{Department of Astronomy and Astrophysics, University of Chicago, Chicago, IL 60637, USA}

\author[0000-0003-4383-2969]{C.~R.~Bom}
\affiliation{Centro Brasileiro de Pesquisas F\'isicas, Rua Dr. Xavier Sigaud 150, 22290-180 Rio de Janeiro, RJ, Brazil}
\author[0000-0002-3690-105X]{J.~A.~Carballo-Bello}
\affiliation{Instituto de Alta Investigaci\'on, Sede Esmeralda, Universidad de Tarapac\'a, Av. Luis Emilio Recabarren 2477, Iquique, Chile}
\author[0000-0002-3936-9628]{J.~L.~Carlin}
\affiliation{Rubin Observatory/AURA, 950 North Cherry Avenue, Tucson, AZ, 85719, USA}
\author[0000-0003-1697-7062]{W.~Cerny}
\affiliation{Kavli Institute for Cosmological Physics, University of Chicago, Chicago, IL 60637, USA}
\affiliation{Department of Astronomy and Astrophysics, University of Chicago, Chicago, IL 60637, USA}
\author[0000-0003-1680-1884]{Y.~Choi}
\affiliation{Department of Astronomy, University of California, Berkeley, Berkeley, CA 94720, USA}
\author[0000-0001-8251-933X]{A.~Drlica-Wagner}
\affiliation{Fermi National Accelerator Laboratory, P.O.\ Box 500, Batavia, IL 60510, USA}
\affiliation{Kavli Institute for Cosmological Physics, University of Chicago, Chicago, IL 60637, USA}
\affiliation{Department of Astronomy and Astrophysics, University of Chicago, Chicago, IL 60637, USA}
\author[0000-0001-5160-4486]{D.~J.~James}
\affiliation{ASTRAVEO LLC, PO Box 1668, Gloucester, MA 01931}
\affiliation{Applied Materials Inc., 35 Dory Road, Gloucester, MA 01930}
\author[0000-0002-9144-7726]{C.~E.~Mart\'inez-V\'azquez}
\affiliation{Gemini Observatory, NSF's NOIRLab, 670 N. A'ohoku Place, Hilo, HI 96720, USA}
\author[0000-0003-0105-9576]{G.~E.~Medina}
\affiliation{Department of Astronomy and Astrophysics, University of Toronto, 50 St. George Street, Toronto ON, M5S 3H4, Canada}
\author[0000-0001-9649-4815]{B.~Mutlu-Pakdil}
\affiliation{Department of Physics and Astronomy, Dartmouth College, Hanover, NH 03755, USA}
\author{M.~Navabi}
\affiliation{Department of Physics, University of Surrey, Guildford GU2 7XH, UK}
\author{N.~E.~D.~No\"el}
\affiliation{Department of Physics, University of Surrey, Guildford GU2 7XH, UK}
\author[0000-0002-1594-1466]{J.~D.~Sakowska}
\affiliation{Department of Physics, University of Surrey, Guildford GU2 7XH, UK}
\author[0000-0003-1479-3059]{G.~S.~Stringfellow}
\affiliation{Center for Astrophysics and Space Astronomy, University of Colorado, 389 UCB, Boulder, CO 80309-0389, USA}
\collaboration{20}{(DELVE Collaboration)}
\correspondingauthor{Mairead Heiger}
\email{mairead.heiger@mail.utoronto.ca}

\begin{abstract}\label{sec:abstract}

We present a spectroscopic analysis of Eridanus IV (Eri~IV) and Centaurus I (Cen~I), two ultra-faint dwarf galaxies of the Milky Way. Using IMACS/Magellan spectroscopy, we identify 28 member stars of Eri~IV and 34 member stars of Cen~I. For Eri~IV, we measure a systemic velocity of $v_{sys} =  -31.5^{+1.3}_{-1.2}\:\mathrm{km\:s^{-1}}$ and velocity dispersion $\sigma_{v}= 6.1^{+1.2}_{-0.9}\:\mathrm{km\:s^{-1}}$. Additionally, we measure the metallicities of 16 member stars of Eri~IV. We find a metallicity of $\mathrm{[Fe/H]}=-2.87^{+0.08}_{-0.07}$ and resolve a dispersion of $\sigma_{\mathrm{[Fe/H]}} = 0.20\pm0.09$. The mean metallicity is marginally lower than all other known ultra-faint dwarf galaxies, making it one of the most metal-poor galaxies discovered thus far. Eri~IV also has a somewhat unusual right-skewed metallicity distribution. For Cen~I, we find a velocity $v_{sys} = 44.9\pm0.8\:\mathrm{km\:s^{-1}}$ and velocity dispersion $\sigma_{v} = 4.2^{+0.6}_{-0.5} \:\mathrm{km\:s^{-1}}$. We measure the metallicities of 27 member stars of Cen~I, and find a mean metallicity $\mathrm{[Fe/H]} = -2.57\pm0.08$ and metallicity dispersion $\sigma_{\mathrm{[Fe/H]}} = 0.38^{+0.07}_{-0.05}$. 
We calculate the systemic proper motion, orbit, and the astrophysical J-factor for each system, the latter of which indicates that Eri~IV is a good target for indirect dark matter detection. We also find no strong evidence for tidal stripping of Cen~I or Eri~IV. Overall, our measurements confirm that Eri~IV and Cen~I are dark matter-dominated galaxies with properties largely consistent with other known ultra-faint dwarf galaxies. The low metallicity, right-skewed metallicity distribution, and high J-factor make Eri~IV an especially interesting candidate for further followup. 

\end{abstract}

\section{Introduction} \label{sec:intro}

Ultra-faint dwarf galaxies (UFDs) are ancient, exceptionally metal-poor dwarf galaxies with low luminosity ($L \lesssim 10^5 L_{\odot}$). Their low baryonic mass, however, belies their dark matter content: they are the most dark matter-dominated objects yet observed (see \citealt{2019ARA&A..57..375S} for a review). Different paradigms of dark matter predict different properties, numbers, and distributions of UFDs, so a complete, well-characterized sample of UFDs can provide important constraints on the nature of dark matter (see \citealt{Bullock2017:2017ARA&A..55..343B} for a review). Additionally, because they form so few stars, stellar feedback is unlikely to have played a large role in altering their density distribution \citep{Penarrubia2012:2012ApJ...759L..42P}. This makes UFDs unique environments for studying the density profiles of the halos in which they reside \citep{2010MNRAS.406.1290P, Orkney2022:2022MNRAS.515..185O}.

In addition to being extremely dark matter-dominated, UFDs are likely among the first galaxies formed. Compared to their larger counterparts, they have simple assembly histories and have undergone little chemical evolution despite their age, so they offer a unique perspective on galaxy formation at high redshift \citep[e.g.,][]{2010Natur.464...72F,2012ApJ...753L..21B,2014ApJ...796...91B, 2015ApJ...805..130K}. Abundance patterns of UFDs also preserve considerable information about enrichment processes and star formation in relatively pristine environments, and inform our understanding of the assembly history of the Milky Way halo \citep[e.g.,][]{1997A&A...326..751N, 2010ApJ...721..738Z, 2012ApJ...759..115F, 2014ApJ...786...74F, 2015ARA&A..53..631F, 2015ApJ...805..130K, 2019ApJ...885..102L}.

The number of UFDs discovered and characterized has risen dramatically in recent years, thanks to dedicated searches of photometric surveys such as the Sloan Digital Sky Survey \citep[e.g.,][]{2007ApJ...654..897B}, the Dark Energy Survey \citep[e.g.,][]{2015ApJ...807...50B}, Pan-STARRS \citep[e.g.,][]{2015ApJ...813...44L, 2020ApJ...893...47D}, and the DECam Local Volume Exploration Survey (DELVE; \citealt{2020ApJ...890..136M,2021ApJ...910...18C, 2022arXiv220912422C,2023ApJ...942..111C}). Proper motion measurements from \textit{Gaia} also facilitate study of the kinematics and orbital properties of these systems, and the discovery of new faint, gravitationally bound systems not yet identified in photometry \citep{2019MNRAS.488.2743T}.

In addition to the wealth of information in photometric and astrometric data, the characterization of new faint stellar systems benefits from spectroscopic followup \citep[e.g.,][]{2007ApJ...670..313S, 2017ApJ...838....8L, 2015ApJ...808...95S, 2020ApJ...892..137S, 2019ApJ...883...11F, 2023ApJ...942..111C, 2023arXiv230203708B}. Radial velocity and metallicity measurements allow us to identify member stars of a given system and provide information on its dark matter content and evolutionary history.

Here, we present follow-up spectroscopy on Eridanus IV (Eri~IV) and Centaurus I (Cen~I), two ultra-faint dwarf galaxy candidates identified using the DELVE survey \citep{2021ApJS..256....2D,2022ApJS..261...38D}. DELVE is a photometric survey of the southern sky dedicated to the study of dwarf galaxies throughout the Local Volume, including faint Milky Way satellites, the Magellanic Clouds, and four isolated analogs of the Magellanic Clouds. To date, twelve new satellites have been identified by DELVE since its first data release in 2021, including Eri~IV and Cen~I \citep{2020ApJ...890..136M, 2021ApJ...910...18C, 2021ApJ...920L..44C, 2022arXiv220912422C, 2023ApJ...942..111C}.

Both Cen~I \citep{2020ApJ...890..136M} and Eri~IV \citep{2021ApJ...920L..44C} were identified using a matched-filter algorithm to search for local spatial overdensities consistent with an old, metal-poor stellar population. We refer to the discovery papers for further details on the method of their discovery. Both systems were identified as ultra-faint dwarf galaxies candidates (rather than star clusters) on the basis of their physical size and absolute magnitude, pending spectroscopic confirmation. Photometry of Eri~IV suggested an age of approximately $13$ Gyr and an upper limit on the metallicity of $\mathrm{[Fe/H]} < -2.08$. Photometry of Cen~I suggested an age of $>12.85$ Gyr and a metallicity of $\mathrm{[Fe/H]} = -1.8$. The morphological parameters determined in the discovery papers are summarized in the first 16 rows of Table \ref{tab:params} \citep{2020ApJ...890..136M,2021ApJ...920L..44C}.

\begin{deluxetable*}{@{}lcllc@{}}
\centering
\tablehead{
\colhead{Parameter}             & \colhead{Description}           & \colhead{Eri IV}                        & \colhead{Cen I}                      & \colhead{Unit}}

\startdata
\midrule
R.A.                 &  Right Ascension (J2000)     & $76.438^{+0.012}_{-0.008}$             & $189.585\pm0.004$                         & deg \\
Decl.                &  Declination (J2000)     & $-9.515^{+0.006}_{-0.007}$             & $-{40.902}_{-0.005}^{+0.004}$             & deg\\
$a_{h}$              &  Angular semi-major axis length     & $4.9^{+1.1}_{-0.8}$                    & ${2.9}_{-0.4}^{+0.5}$                     & arcmin \\
$r_{h}$              &  Azimuthally-averaged angular half-light radius     & $3.3^{+0.7}_{-0.6}$                    & ${2.3}_{-0.3}^{+0.4}$                     & arcmin  \\
$r_{1/2}$            &  Azimuthally-averaged physical half-light radius     & $75^{+16}_{-13}$                       & ${79}_{-10}^{+14}$                        & pc  \\
$\varepsilon$        &  Ellipticity     & $0.54^{+0.10}_{-0.14}$                 & ${0.4}\pm0.1$                     & -- \\
P.A.                 &  Position angle of major axis (East of North)     & $65^{+9}_{-8}$                         & ${20}\pm11$                        & deg \\
$m$                  &  Distance modulus      & $19.42^{+ 0.01}_{-0.08} \pm 0.1$       & ${20.354}\pm {-0.072}$                    & mag\\
$D_{\odot}$          &  Heliocentric distance      & $76.7^{+4.0}_{-6.1}$                   & ${117.7}\pm 4.1$                          & kpc\\
$\tau$               &  Age     & $13.0^{+0.1}_{-1.0}$                   & $>$12.85                                  & Gyr \\
$M_V$                &  Absolute (integrated) V-band magnitude     & $-4.7 \pm 0.2$                         & $-{5.55}\pm0.11$                 & mag  \\
$\mu$                &  Average surface brightness within the half-light radius    & 28.2                                   & 27.9                                      & mag~arcsec$^{-2}$\\
$M_{*}$              &  Stellar mass     & $6519^{+1038}_{-1111}$                 & $14300\pm1800$                            & $\mathrm{M_{\odot}}$ \\
$l$                  &  Galactic longitude     & 209.499                                & 300.265                                   & deg \\
$b$                  &  Galactic latitude     & -27.772                                & 21.902                                    & deg \\
$E(B-V)$             &  Mean reddening within the half-light radius      & 0.107                                  & 0.124                                     & mag \\
\midrule\midrule
$v_{sys}$                   & Systemic radial velocity in the heliocentric frame    & $-31.5^{+1.3}_{-1.2}$   & $44.9\pm0.8$                                 & $\mathrm{km\:s}^{-1}$\\
$\sigma_{v}$          & Velocity dispersion & $6.1^{+1.2}_{-0.9}$             & $4.2^{+0.6}_{-0.5}$                          & $\mathrm{km\:s}^{-1}$ \\
$M_{1/2}$                   & Enclosed mass within one half-light radius            & $1.90^{+1.05}_{-0.70}\times10^6$ & $1.36^{+0.39}_{-0.28}\times10^{6}$ & $M_{\odot}$ \\
$M_{1/2}/L_{1/2}$                 & Mass-to-light ratio & $598^{+361}_{-228}$        & $140^{+60}_{-42}$                         & $M_{\odot} L_{\odot}^{-1}$ \\
$\mathrm{[Fe/H]}$           & Mean metallicity & $-2.87^{+0.08}_{-0.07}$    & $-2.57\pm0.08$                            & dex \\
$\sigma_{[\mathrm{Fe/H]}}$  & Metallicity dispersion & $0.20\pm0.09$              & $0.38^{+0.07}_{-0.05}$                    & dex \\
$\mu_{\alpha} \cos \delta$  & Systemic proper motion in R.A. & $0.22\pm 0.06$            & $-0.14\pm0.05$                             & mas~yr$^{-1}$  \\
$\mu_{\delta}$              & Systemic proper motion in Decl. & $-0.11\pm 0.05$           & $-0.19\pm0.04$                             & mas~yr$^{-1}$ \\ 
\midrule
$r_{apo}$                   & Orbital apocenter & $135^{+24}_{-13}$   & $212^{+39}_{-17}$          & kpc \\
$r_{peri}$                  & Orbital pericenter & $43\pm11$  & $32^{+12}_{-8}$            & kpc \\
$e$                         & Orbital eccentricity & $0.52^{+0.07}_{-0.03}$  & $0.74^{+0.05}_{-0.04}$           & -- \\ 
\midrule
$\log_{10}{J(0.2^{\circ})}$ & Integrated J-factor within a solid angle of 0.2° & $18.6^{+0.4}_{-0.3}$ & $17.7\pm0.3$ & ${\rm GeV^2~cm^{-5}}$\\ 
$\log_{10}{J(0.5^{\circ})}$ & Integrated J-factor within a solid angle of 0.5° & $18.8\pm0.4$ & $17.7_{-0.3}^{+0.4}$  & ${\rm GeV^2~cm^{-5}}$\\
\bottomrule
\enddata
\caption{Morphological, kinematic, chemical, and orbital properties of Eri~IV and Cen~I. Parameters R.A. through $E(B-V)$ are from \citet{2021ApJ...920L..44C} and \citet{2020ApJ...890..136M}; $v_{sys}$ through $\log_{10}{J(0.5^{\circ})}$ are calculated in this work.\label{tab:params}}
\end{deluxetable*}

This work is organized as follows: Section \ref{sec:data} outlines the observations and data, including the target selection process and data reduction. Section \ref{sec:meas} details the method used to measure the velocities and metallicities of observed stars. Section \ref{sec:mems} addresses the identification and nature of the member stars. The systemic properties such as the velocity dispersion and mean metallicity are presented in Section \ref{sec:sys}. In Section \ref{sec:discussion}, we calculate the mass-to-light ratio of Eri~IV and Cen~I, model their orbits, investigate the possibility of tidal stripping, address the metallicity distribution function of Eri~IV specifically, and discuss the properties of both galaxies in the context of the current census of Milky Way satellites. We summarize our results in Section \ref{sec:summary}.

\section{Observation and Data}\label{sec:data}

We observed Eri~IV and Cen~I with the Inamori-Magellan Areal Camera and Spectrograph (IMACS) on the 6.5m Magellan-Baade telescope \citep{2011PASP..123..288D}, each with three multi-slit masks. All observations used the instrument's f/4 camera and 1200 lines/mm grating blazed at $9000$ \AA, with a spectral resolution of $\lambda/\Delta\lambda \approx 11,000$. The spectra are centered at $8200$ \AA\ and span roughly $7400 - 9000$ \AA. This wavelength range is chosen to cover the infrared Ca II triplet (CaT) absorption lines at $8498$ \AA, $8542$ \AA\ and $8662$ \AA, and atmospheric absorption (telluric) features at $7600$ \AA. 

Eri~IV was observed using three multi-slit masks (denoted as Eri4\_mask1, Eri4\_mask2, and Eri4\_mask3) during two separate observing runs: first on October 9th, 2021, and again during a three-night observing run between January 28th and January 30th, 2022. He, Kr, Ar, and Ne arc lamp calibration frames and flat frames were taken every 1-3 science exposures (similar to the procedure in \citealt{2017ApJ...838....8L}). The details of these observations are presented in Table \ref{tab:obs}. In total, 80 unique stars were observed using these masks and 124 total spectra were collected. The spectra taken with the same mask during the same run are co-added. The different runs are kept separate for purposes of identifying variability. 

Cen~I was observed using three multi-slit masks (denoted as Cen1\_mask1, Cen1\_mask2, and Cen1\_mask3) during three separate observing runs: February 10th, 2020, July 11th-13th 2021, and the same three-night observing run in January 2022 as Eri~IV. The calibration procedure was the same as for Eri~IV, and the spectra are co-added by mask within a run. In total, 228 spectra of 110 unique stars were collected. See Table \ref{tab:obs} for more detail on the observations.

We selected the targets from three categories. First, we selected all candidates with high membership probability from the proper motion mixture models presented in \citet{2021ApJ...920L..44C, 2022ApJ...940..136P} as the highest priority targets. The mixture model methodology is described in detail in \citet{2022ApJ...940..136P}. Briefly, we construct a mixture model of a dwarf and Milky Way foreground components and utilize the \textit{Gaia} proper motions \citep{GaiaCollaboration2021:2021A&A...649A...1G,Lindegren2021:2021A&A...649A...2L} and spatial positions to differentiate the components. The model includes a color-magnitude filter using an old, metal-poor isochrone with DELVE DR2 photometry to further remove MW foreground. 

We then prioritized possible member candidates present in \textit{Gaia} but not in DELVE (due to their proximity to a bright object) whose proper motions are consistent with the UFDs. Finally, as the lowest priority, we included fainter candidates that are not in \textit{Gaia} and only in DELVE DR2, guided by an old, metal-poor isochrone. These targets have no \textit{Gaia} \texttt{source\_id} and therefore are identified by their DELVE \texttt{QUICK\_OBJECT\_ID}. We designed three slitmasks for each dwarf galaxy. With the exception of Cen1\_mask1, which was based on targets selected from \textit{Gaia} DR2 in 2020, all other masks are designed using targets selected from \textit{Gaia} EDR3.

The IMACS data were reduced following the procedure outlined in \citet{2017ApJ...838...11S}. First, a bias subtraction was performed. A preliminary wavelength solution and mapping of each slit were then applied using the Cosmos pipeline \citep{2011PASP..123..288D, 2017ascl.soft05001O}, followed by an adaptation of the DEIMOS DEEP2 pipeline \citep{2012ascl.soft03003C, 2013ApJS..208....5N} for the remainder of the data reduction, including a flat-field correction, sky subtraction, and final spectral extraction.

We use photometric data from DELVE DR2 \citep{2022ApJS..261...38D} and \textit{Gaia} EDR3 \citep{GaiaCollaboration2021:2021A&A...649A...1G,2021A&A...649A...3R}. A reddening correction is applied to DELVE DR2 $gri$ photometry using the interstellar extinction values provided by the DELVE DR2 catalogue \citep{2022ApJS..261...38D}. For \textit{Gaia} photometry, we use the color excess $E(B-V)$ from \citet{Schlegel1998:1998ApJ...500..525S} (making use of the \texttt{dustmap} module by \citealt{Green2018:2018JOSS....3..695G}) and the color-dependent relations from \citet{GaiaCollaboration2018:2018A&A...616A..10G} to compute the extinction.

\begin{deluxetable*}{@{}lllllcrcccc@{}}
\tablecaption{Summary of IMACS observations.\label{tab:obs}}

\tablehead{
\colhead{MJD$^{a}$} & \colhead{Mask}  & \colhead{$\alpha(J2000)$} & \colhead{$\delta(J2000)$} & \colhead{Slit PA}       & \colhead{\# of}   &\colhead{$\sum{t_{exp}}$} & \colhead{Seeing} & \colhead{\# objects} & \colhead{\# useful}  \\
              &                 &\colhead{(h:m:s)}         & \colhead{(\degr \arcmin \arcsec )} & \colhead{(\degr)} & \colhead{exp.} &   \colhead{(s)}   & \colhead{\arcsec}   & \colhead{observed}   &  \colhead{spectra$^{b}$}}
\startdata
59497.3 & Eri4\_mask1 & 05:05:49.200    & -09:29:20.00   & 32.52    & 4      &  7720           & 1.00    & 45          & 10   \\
59608.6 & Eri4\_mask1 & 05:05:49.200    & -09:29:20.00   & 32.52    & 9      & 16200           & 0.60    & 45          & 31   \\ 
59609.2 & Eri4\_mask2 & 05:05:35.000    & -09:35:35.00   & 32.50    & 3      &  3840           & 0.70    & 26          & 15   \\
59610.1 & Eri4\_mask3 & 05:06:09.454    & -09:31:07.20   & 32.45    & 3      &  5400           & 0.67    & 22          & 12   \\ 
58881.3 & Cen1\_mask1 & 12:38:22.940    & -40:54:36.00   & 32.57    & 1      &  7200           & 0.80    & 64          & 32   \\
59410.0 & Cen1\_mask2 & 12:38:21.987    & -40:57:07.00   & 32.54    & 4      &  9560           & 1.50    & 57          & 16   \\
59608.6 & Cen1\_mask2 & 12:38:21.987    & -40:57:07.00   & 32.31    & 6      & 25620           & 0.50    & 57          & 42   \\
59609.3 & Cen1\_mask3 & 12:38:34.688    & -40:52:19.19   & 32.50    & 3      &  6330           & 0.50    & 50          & 18 \\
\enddata  
\tablenotetext{a}{The MJD presented is the average over multiple nights of a single run (where applicable).}
\tablenotetext{b}{The number of observed spectra with reliable radial velocity fits.}
\end{deluxetable*}

\section{Radial velocity and metallicity measurements}\label{sec:meas}

\subsection{Radial velocity}\label{sec:rv}
The radial velocity of each star is measured using a technique detailed in \citet{2017ApJ...838....8L}. One of three template spectra (HD122563, a metal-poor red giant branch (RGB) star; HD26297, a somewhat more metal-rich RGB star; or HD161817, a blue horizontal-branch (BHB) star) is applied to the reduced observed spectrum. The best fit velocity, $v_{obs}$, and the best-fit template are those which maximize the likelihood function: 
\begin{equation}
    \log{\mathcal{L}}=-\frac{1}{2} \sum_{\lambda = \lambda_1}^{\lambda_2} \frac{[f_s(\lambda) - f_{std}(\lambda(1+\frac{v_{obs}}{c}))]^2}{\sigma^2_s(\lambda)}
    \label{eqn:like}
\end{equation}

In the likelihood function, $f_s(\lambda)$ and $\sigma_s^2(\lambda)$ are the normalized spectrum and its variance and $f_{std}(\lambda)$ is the normalized template spectrum. A star may have multiple reduced spectra from the same epoch and/or multiple reduced spectra from different epochs, so this technique is applied to each such spectrum individually. The RGB templates fit the CaT lines over a wavelength range of $8450 - 8685$ \AA, and the BHB template fits the Paschen lines over a wider range, from $8450$ \AA\  to $8900$ \AA. The templates were obtained during previous IMACS observations and were reduced with the same process as these data \citep{2017ApJ...838....8L, 2018ApJ...857..145L}. 

For each spectrum, we used the Markov Chain Monte Carlo (MCMC) sampler \texttt{emcee} \citep{2013PASP..125..306F} to sample the likelihood function (Equation \ref{eqn:like}), using 20 walkers, each taking 1000 steps, with a burn-in period of 50 steps. The chains converge quickly (the starting position is determined with a maximum likelihood estimate), so longer sampling or burn-in is not necessary. The line-of-sight velocity $v_{obs}$ and its uncertainty, $\sigma_{v_{obs}}$, are taken to be the median velocity and standard deviation of the posterior distributions. These results were also inspected for visual agreement between the template fit and the observed line profile and to verify convergence.

A telluric correction $v_{tell}$, which accounts for miscentering of stars within slits, is determined by the same process. A telluric template (also obtained during previous IMACS observations) is fit to the atmospheric absorption lines in the wavelength range $\lambda = 7550 - 7700 $ \AA, and the median and standard deviation give $v_{tell}$ and $\sigma_{v_{tell}}$. The telluric correction and a time-dependent heliocentric transformation are then applied to the observed velocity to give the heliocentric velocity $v_{hel}$. 

To estimate the uncertainty $\sigma_{v_{hel}}$, statistical uncertainties that arise from the fitting procedure and a systematic error of $1\:\mathrm{km\:s^{-1}}$ are added in quadrature, i.e. $\sigma_{v_{hel}} = \sqrt{\sigma_{v_{obs}}^2 + \sigma_{v_{tell}}^2 + \sigma_{sys}^2}$. The systematic error, which accounts for uncertainties like calibration uncertainties or template mismatching, is based on previous work on the repeatability of IMACS observations of UFDs \citep{2017ApJ...838....8L}. The dominant source of uncertainty is typically the systematic uncertainty except at low S/N, at which point $\sigma_{v_{obs}}$ dominates.

If there are multiple velocity measurements of an object, the inverse variance weighted average of the measurements and the standard error of the weighted mean are used as $v_{hel}$ and $\sigma_{v_{hel}}$ when calculating systemic properties. The measurements in Table \ref{tab:eri4mems} and Table \ref{tab:cen1mems} show the individual measurements rather than these averages. We also use the independent measurements of radial velocity from different runs to look for evidence of binary stars. 

\subsection{Metallicity}\label{sec:met}
To measure the metallicity, we use the equivalent widths (EWs) of the CaT lines. These absorption features are prominent in the near-infrared spectra of RGB stars, so they can be measured with medium resolution spectroscopy even at these faint magnitudes \citep{2013MNRAS.434.1681C}. Additionally, an empirical relationship between the EW of these lines and metallicity has been observed and utilized for decades \citep[e.g.,][]{1997PASP..109..907R}. We employ the calibration from \citet{2013MNRAS.434.1681C} to make use of this relationship and convert the EW of the CaT lines to metallicity. The calibration is only valid for RGB stars, so horizontal branch (HB) stars are excluded.

To determine the EW, each line is fit using with a Gaussian + Lorentzian line profile using \texttt{mpfit}, which performs a non-linear least squares fit \citep{2009ASPC..411..251M}. The three EWs are summed and converted to metallicity per \citet{2013MNRAS.434.1681C}, as in \citet{2017ApJ...838....8L} and \citet{2015ApJ...808...95S}. For stars with multiple EW measurements, the inverse variance weighted average of the EW is used to calculate the metallicity. All epochs and masks are included in the average.

Occasionally, there are issues with the fit that affect only one of the three lines in the observed spectrum (for example, a chip gap). For these stars, we take advantage of the proportionality between the three lines and use the EWs of the two usable lines to estimate the EW of the missing line. We performed orthogonal distance regression on a dataset of 555 stars in the dwarf galaxies Carina II and Carina III observed with IMACS \citep{2018ApJ...857..145L}, plus member stars in Cen~I and Eri~IV with usable EW measurements from \texttt{mpfit}, to obtain the following linear relationships: 

\begin{equation}
    EW1 = 0.41\times EW2 + 0.14
\end{equation} 
\begin{equation}
    EW1 = 0.56\times EW3 + 0.06
\end{equation}
\begin{equation}
    EW2 = 2.41\times EW1 - 0.33
\end{equation}
\begin{equation}
    EW2 = 1.35\times EW3 - 0.21    
\end{equation}
\begin{equation}
    EW3 = 1.80\times EW1 - 0.10
\end{equation}
\begin{equation}
    EW3 = 0.74\times EW2 + 0.16    
\end{equation}
    
The two predictions for each line are summed in quadrature, and the uncertainty is the sum in quadrature of the uncertainty in the EW of the two lines used to predict the third and uncertainty in the regression parameters. 

We validate these relationships first by comparing the predicted and measured EW of each line for all stars in the dataset, and find that the standard deviation of the residuals is $0.18$, $0.43$, and $0.34$ \AA\ for EW1, EW2, and EW3 respectively. Additionally, for Cen~I and Eri~IV members, we replaced the measured EW of one line with the predicted EW and re-calculated the metallicity. For Cen~I, the standard deviation of the residuals of the metallicity is $0.07$, $0.09$, and $0.07$ dex when replacing EW1, EW2, and EW3 respectively. For Eri~IV, the standard deviation of the residuals of the metallicity is $0.10$, $0.13$, and $0.11$ dex. In case the proportionality is metallicity dependent, this regression was repeated on a subsample of only those stars with comparable EW to Eri~IV and Cen~I (213 stars, including Eri~IV and Cen~I members), but there was no meaningful difference in the parameters, residuals, or predicted metallicity when fitting only to more metal-poor stars.

We use the relations derived to predict the equivalent width and then metallicity for the three stars in Cen~I, noted in Table \ref{tab:cen1mems}, that lack a measurement for one of the CaT lines (two lack EW1 and one lacks EW3). These relations are not used for any members of Eri~IV.

To remove dependencies of the EW on temperature and surface gravity, the calibration uses the absolute V-band magnitude of the object in addition to the sum of the EWs. We convert from \textit{Gaia} passbands per \citet{2021A&A...649A...3R} and DELVE $gr$-bands per \citet{2015ApJ...807...50B} to absolute V-band magnitude. We do not have complete coverage of either Cen~I or Eri~IV in one single photometric system (see the cyan/magenta outlined stars in Figure \ref{fig:summary}), so DELVE photometry is used by default and \textit{Gaia} photometry when necessary. DELVE photometry is preferred because it is more accurate at fainter magnitudes and is used to distinguish RGB/HB stars. To ensure that the choice of photometric system does not unduly influence the metallicity, we compared the absolute V-band magnitude and metallicity for 13 stars in Eri~IV present in both catalogues and found no evident bias. The resulting difference in metallicity is $<0.02$ dex, and most differ by $<0.01\:\mathrm{dex}$. 

The reported uncertainties include uncertainties on the summed EW, the distance, the photometry, and the calibration coefficients. The uncertainty of the summed EW measurements is the sum in quadrature of a $0.2$ \AA\ systematic floor \citep{2017ApJ...838....8L} and the statistical error from the fitting process, which is the sum in quadrature of the uncertainty associated with each line.

\section{Spectroscopic Membership} \label{sec:mems}

\subsection{Eridanus~IV}\label{sec:eri4mems}
\begin{figure*}\centering
\includegraphics[width = 1\textwidth]{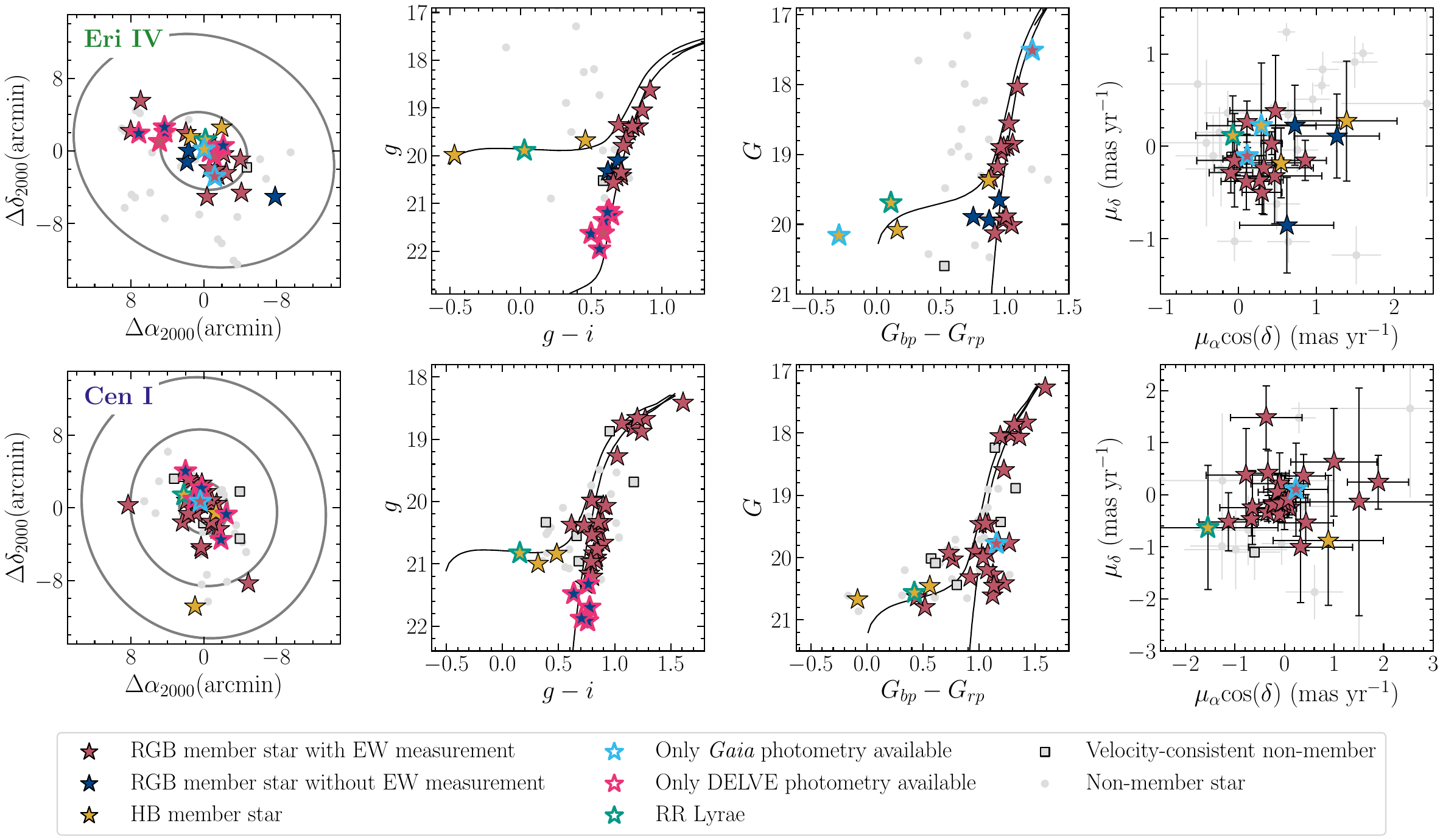}\hfil
\caption{Diagnostic plots of observed stars near Eri~IV (top) and Cen~I (bottom) with reliable radial velocity measurements. Objects outlined in green are RR Lyrae, objects outlined in cyan are available only in \textit{Gaia}, and objects outlined in magenta are available only in DELVE. (Left) Spatial distribution overlaid with ellipses marking 1 and 3 half-light radii of Eri~IV and 1, 3, and 5 half-light radii of Cen~I. All member stars of Eri~IV lie within 3 half-light radii and all member stars of Cen~I lie within 5 half-light radii. (Center left) Color-magnitude diagram in DELVE $gi$-bands with $13\:\mathrm{Gyr}$, $\mathrm{[Fe/H]} = - 2.2$ PARSEC isochrone overplotted. (Center right) Color-magnitude diagram in \textit{Gaia} $G-$band and $G_{bp}-G_{rp}$, overlaid again with the PARSEC isochrone. (Right) Proper motion. The proper motion of several velocity-consistent non-members of Eri~IV and Cen~I are beyond the range of this plot.
\label{fig:summary}}
\end{figure*}

\begin{figure*}\centering
\includegraphics[width =0.9\textwidth]{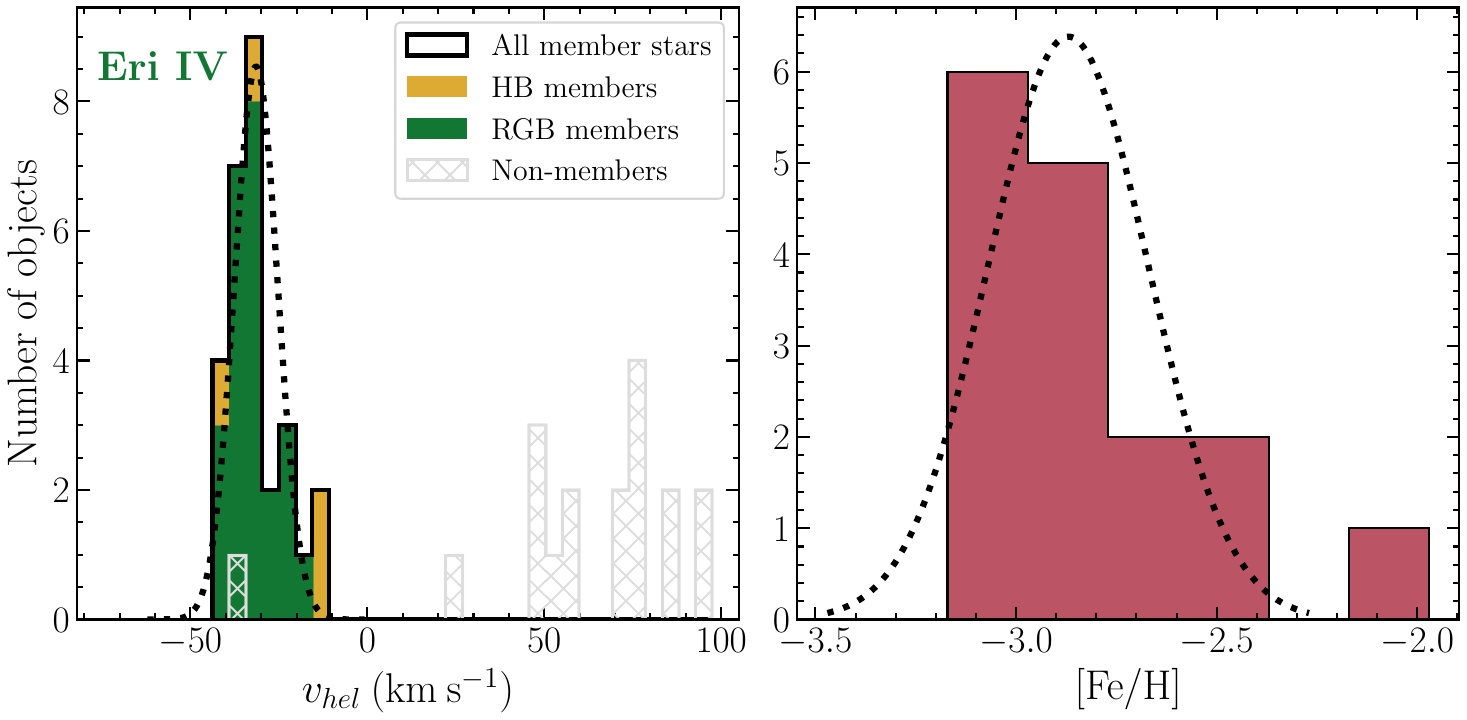}
\caption{(Left) Distribution in velocity of the observed stars near Eri~IV with reliable radial velocity measurements. A Gaussian model of the systemic velocity and dispersion is overplotted in a black dotted line. The total sample of members is outlined in black, and a stacked histogram of members according to evolutionary stage is plotted in green (RGB stars) and yellow (HB stars). Non-members are cross-hatched. The range is chosen to highlight the member stars and the contrast between member and non-member stars. It excludes 7 non-member stars whose velocities exceed $100\: \mathrm{km\:s^{-1}}$. (Right) Distribution of $\mathrm{[Fe/H]}$ of 16 RGB member stars of Eri~IV, overplotted with a black dotted line showing the Gaussian model of the systemic metallicity and dispersion. 
\label{fig:eri4hist}}
\end{figure*}

Of the 80 stars observed in the field of Eri~IV, visual inspection of the spectra showed that 53 have reliable radial velocity fits, and 11 of these have more than one reliable radial velocity fit from repeat observations. We identify 28 member stars of Eri~IV in this 53 star sample (8 of which have multiple measurements) that are tightly grouped in radial velocity between $-50\:\mathrm{km\:s^{-1}}$ and $0\:\mathrm{km\:s^{-1}}$, as seen in the left panel of Figure \ref{fig:eri4hist}. We consider a star to be a member if its velocity is $-50 \: \mathrm{km\:s^{-1}} < v_{hel} < 0 \: \mathrm{km\:s^{-1}}$, provided it is within five half-light radii, its proper motion is consistent with other likely members (within $3\sigma$ of the mean proper motion), and it lies along the isochrone. These criteria yield member stars that are all very metal-poor, so there was no specific metallicity criterion. All non-member stars but one are excluded on the basis of velocity, which is excluded on the basis of proper motion. The proper motion of this candidate star (\texttt{source\_id}: 3182723151776676352) is more than $5\sigma$ from the mean---it is not visible in the proper motion panel of Figure \ref{fig:summary} as it lies outside of the range of the plot at ($5.3\pm2.8$, $4.5\pm1.8$) $\mathrm{mas\:yr^{-1}}$. As such, it is highly unlikely to be a member star and we consider it a field star.

The top panel of Figure \ref{fig:summary} summarizes the properties of the member stars, as well as observed non-members with reliable radial velocity measurements. Members are denoted by stars in yellow (HB), red (RGB with metallicity measurement), and navy (RGB without metallicity measurement). Non-members are denoted by grey squares (non-member stars with velocity consistent with Eri~IV) and grey circles (non-member stars with velocity inconsistent with Eri~IV). The leftmost panel shows that all member stars of Eri~IV lie within three half-light radii of the centroid. Some observed stars lie within five half-light radii, but were determined to be non-members. The member stars are also situated along a $13\:\mathrm{Gyr}$, $\mathrm{[Fe/H]} = - 2.2$ (the most metal-poor available) PARSEC isochrone \citep{2012MNRAS.427..127B,2017ApJ...835...77M}, as seen in the two center panels. Lastly, the rightmost panel shows that the stars are clustered in proper motion space. Seven member stars are present only in DELVE, and two are only in \textit{Gaia}. They are outlined in Figure \ref{fig:summary} in magenta and cyan respectively.

We differentiate four stars that lie along the horizontal branch; the remaining 24 are RGB stars. As the calibration used to measure metallicity is valid only for RGB stars, and because reliable EW measurements require a higher S/N than radial velocity measurements, we successfully measured the metallicity of 16 of the 24 RGB member stars, presented in the right panel of Figure \ref{fig:eri4hist}. Table \ref{tab:eri4mems} presents the details of the 28 member stars of Eri~IV. 

\begin{deluxetable*}{rrrrrrrcrrrrc}
\tabletypesize{\footnotesize}
\setlength{\tabcolsep}{4pt}
\tablecaption{Spectroscopically confirmed members of Eri~IV.
\label{tab:eri4mems}}
\tablehead{
\colhead{ID} & \colhead{R.A.}  & \colhead{Decl.} & \colhead{$g_0$} & \colhead{$r_0$} & \colhead{$G$}   &\colhead{MJD} & \colhead{Mask} & \colhead{S/N} & \colhead{$v_{hel}$} & \colhead{EW} & \colhead{[Fe/H]}  & \colhead{Notes} \\
           & \colhead{(deg)} &\colhead{(deg)} & \colhead{(mag)} & \colhead{(mag)} & \colhead{(mag)} &            &             &             & \colhead{(km/s)}  & \colhead{(\AA)} & \colhead{(dex)} & }
\startdata
3182749196457569664 & 76.556175 & -9.423492 & 20.56 & 20.11 & 20.12 & 59608.6 & 1 & 8.1 & -34.7$\pm$1.6 & 1.96$\pm$0.68 & -2.66$\pm$0.30 & RGB \\ 
3182747753348514816 & 76.574429 & -9.478353 & 20.35 & 19.82 & 19.88 & 59608.6 & 1 & 13.7 & -32.8$\pm$1.2 & 2.39$\pm$0.55 & -2.48$\pm$0.21 & RGB, BIN \\ 
... & ... & ... & ... & ... & ... & 59497.3 & 1 & 4.1 & -39.4$\pm$2.7 & 2.29$\pm$1.38 & -2.53$\pm$0.63 & ... \\ 
... & ... & ... & ... & ... & ... & 59610.1 & 3 & 3.9 & -30.1$\pm$3.2 & -- & -- & ... \\ 
3182736758232241280 & 76.405171 & -9.472447 & 19.99 & 20.25 & 20.08 & 59608.6 & 1 & 6.2 & -15.1$\pm$5.5 & -- & -- & HB \\ 
3182725179000404992 & 76.471771 & -9.481678 & 19.79 & 19.32 & 19.33 & 59608.6 & 1 & 23.1 & -37.2$\pm$1.1 & 1.77$\pm$0.44 & -2.93$\pm$0.20 & RGB \\ 
3182725075921186432 & 76.462933 & -9.489581 & 19.69 & 19.36 & 19.37 & 59608.6 & 1 & 20.2 & -32.6$\pm$1.1 & -- & -- & HB \\ 
3182724972841956736 & 76.466850 & -9.514464 & 20.10 & 19.64 & 19.65 & 59610.1 & 3 & 3.7 & -30.4$\pm$2.9 & -- & -- & RGB \\ 
3182724869762754048 & 76.435904 & -9.494575 & 19.90 & 19.90 & 19.69 & 59608.6 & 1 & 8.3 & -43.7$\pm$2.1 & -- & -- & HB, RRL \\ 
3182724869762752640 & 76.433521 & -9.496436 & 19.66 & 19.14 & 19.18 & 59608.6 & 1 & 25.0 & -34.6$\pm$1.1 & 1.67$\pm$0.54 & -3.02$\pm$0.26 & RGB \\ 
... & ... & ... & ... & ... & ... & 59497.3 & 1 & 8.3 & -33.6$\pm$1.6 & 1.63$\pm$0.59 & -3.04$\pm$0.29 & ... \\ 
3182724835403011584 & 76.420108 & -9.503208 & 19.39 & 18.84 & 18.89 & 59608.6 & 1 & 30.3 & -29.2$\pm$1.0 & 1.55$\pm$0.41 & -3.15$\pm$0.21 & RGB \\ 
... & ... & ... & ... & ... & ... & 59497.3 & 1 & 10.9 & -29.4$\pm$1.5 & 2.07$\pm$0.35 & -2.84$\pm$0.14 & ... \\ 
3182724732324650624 & 76.401617 & -9.520436 & 20.44 & 19.95 & 20.00 & 59608.6 & 1 & 12.8 & -34.0$\pm$1.2 & 2.57$\pm$0.81 & -2.37$\pm$0.32 & RGB \\ 
... & ... & ... & ... & ... & ... & 59497.3 & 1 & 3.4 & -37.2$\pm$3.4 & 2.15$\pm$1.15 & -2.58$\pm$0.53 & ... \\ 
3182724663604313600 & 76.437958 & -9.512189 & -- & -- & 20.16 & 59608.6 & 1 & 6.0 & -11.7$\pm$6.7 & -- & -- & HB \\ 
3182724427381796480 & 76.428617 & -9.546403 & 20.42 & 19.93 & 19.94 & 59608.6 & 1 & 13.1 & -20.1$\pm$1.2 & 3.31$\pm$0.61 & -2.02$\pm$0.22 & RGB \\ 
3182724423086994048 & 76.416629 & -9.548908 & 19.36 & 18.86 & 18.90 & 59608.6 & 1 & 27.7 & -39.4$\pm$1.1 & 1.79$\pm$0.24 & -3.00$\pm$0.14 & RGB \\ 
... & ... & ... & ... & ... & ... & 59497.3 & 1 & 11.0 & -39.5$\pm$1.4 & 1.64$\pm$0.33 & -3.09$\pm$0.17 & ... \\ 
3182723770251103232 & 76.470017 & -9.534639 & 20.32 & 19.88 & 19.89 & 59610.1 & 3 & 3.3 & -39.2$\pm$2.8 & -- & -- & RGB \\ 
3182723289215635712 & 76.370492 & -9.530297 & 19.48 & 18.93 & 18.97 & 59608.6 & 1 & 24.5 & -40.7$\pm$1.1 & 1.67$\pm$0.50 & -3.06$\pm$0.24 & RGB \\ 
... & ... & ... & ... & ... & ... & 59497.3 & 1 & 10.7 & -42.1$\pm$1.4 & 1.77$\pm$0.55 & -3.00$\pm$0.25 & ... \\ 
... & ... & ... & ... & ... & ... & 59609.2 & 2 & 9.0 & -38.6$\pm$1.6 & 1.90$\pm$0.76 & -2.92$\pm$0.34 & ... \\ 
3182722979977980544 & 76.396754 & -9.554347 & 18.63 & 17.98 & 18.04 & 59608.6 & 1 & 47.2 & -35.8$\pm$1.0 & 2.00$\pm$0.32 & -3.04$\pm$0.14 & RGB \\ 
... & ... & ... & ... & ... & ... & 59497.3 & 1 & 23.2 & -34.9$\pm$1.1 & 2.09$\pm$0.38 & -3.00$\pm$0.15 & ... \\ 
... & ... & ... & ... & ... & ... & 59609.2 & 2 & 19.1 & -33.6$\pm$1.1 & 2.17$\pm$0.26 & -2.95$\pm$0.12 & ... \\ 
3182722945618240128 & 76.412567 & -9.559406 & 19.38 & 18.8 & 18.85 & 59609.2 & 2 & 10.2 & -32.5$\pm$1.3 & 2.18$\pm$0.58 & -2.79$\pm$0.23 & RGB \\ 
3182722567661099520 & 76.305854 & -9.598422 & 20.37 & 19.92 & 19.94 & 59609.2 & 2 & 3.7 & -38.9$\pm$2.6 & -- & -- & RGB \\ 
3182722395862411776 & 76.368508 & -9.590911 & 19.40 & 18.84 & 18.89 & 59609.2 & 2 & 10.2 & -35.2$\pm$1.4 & 1.9$\pm$0.92 & -2.94$\pm$0.43 & RGB \\ 
3182722194000359808 & 76.418325 & -9.562303 & -- & -- & 17.51 & 59608.6 & 1 & 68.4 & -34.7$\pm$1.0 & 1.82$\pm$0.31 & -3.23$\pm$0.14 & RGB \\ 
... & ... & ... & -- & -- & ... & 59497.3 & 1 & 34.8 & -33.8$\pm$1.0 & 2.11$\pm$0.45 & -3.08$\pm$0.17 & ... \\ 
3182721884761205888 & 76.432442 & -9.598456 & 19.05 & 18.38 & 18.55 & 59609.2 & 2 & 13.7 & -23.2$\pm$1.2 & 2.07$\pm$0.55 & -2.92$\pm$0.22 & RGB \\ 
10713101074373 & 76.511158 & -9.47165 & 21.95 & 21.52 & -- & 59608.6 & 1 & 3.3 & -34.9$\pm$3.0 & -- & -- & RGB \\ 
10713101074247 & 76.402321 & -9.505528 & 21.18 & 20.70 & -- & 59608.6 & 1 & 5.6 & -23.3$\pm$2.0 & -- & -- & RGB \\ 
10713101061181 & 76.419392 & -9.524417 & 21.61 & 21.12 & -- & 59608.6 & 1 & 4.1 & -31.7$\pm$2.5 & 1.58$\pm$0.50 & -2.70$\pm$0.26 & RGB \\ 
10713100987698 & 76.428329 & -9.517106 & 21.36 & 20.99 & -- & 59608.6 & 1 & 5.4 & -34.9$\pm$2.3 & -- & -- & RGB \\ 
10713100010003 & 76.558817 & -9.482972 & 21.25 & 20.78 & -- & 59608.6 & 1 & 5.8 & -26.0$\pm$2.0 & -- & -- & RGB \\ 
10713100009957 & 76.520308 & -9.498281 & 21.15 & 20.67 & -- & 59608.6 & 1 & 7.4 & -29.8$\pm$1.7 & 1.21$\pm$0.68 & -3.08$\pm$0.49 & RGB \\ 
10713100009950 & 76.509917 & -9.481247 & 21.62 & 21.19 & -- & 59608.6 & 1 & 4.1 & -21.3$\pm$2.8 & -- & -- & RGB 
\enddata
\end{deluxetable*}

The 28 member stars include one RR Lyrae (flagged in Table \ref{tab:eri4mems} as RRL and outlined in Figure \ref{fig:summary} in green), which is found in the PS1 variability catalogue from \citet{2017AJ....153..204S} (as noted in the discovery paper). We exclude this star from any calculations involving radial velocity. Searches of the \textit{Gaia} DR3 variability catalogue did not yield any additional RR Lyrae variables amongst our member stars \citep{2022arXiv220606278C}.

To identify possible binaries in our sample, we compared the radial velocity measurements from spectra taken in October 2021 and January 2022 for the 8 members with repeat observations and find one likely binary (flagged in Table \ref{tab:eri4mems} as BIN). The measured velocity from October 2021 using Eri4\_mask1 (MJD 59497.3) is $ v_{hel} = -39.4\pm2.7\:\mathrm{km\:s^{-1}}$, versus $-32.8\pm1.2\:\mathrm{km\:s^{-1}}$ from January 2022 using Eri4\_mask1 (MJD 59608.6) and $-30.1\pm3.2\:\mathrm{km\:s^{-1}}$ using Eri4\_mask3 (MJD 59610.1). The differences between the measurement made in October 2021 and the measurements made in January 2022 are $|\Delta v_{hel}| = 6.6\pm3.0\:\mathrm{km\:s^{-1}}$ (Eri4\_mask1, MJD 59608.6) and $|\Delta v_{hel}| = 9.3\pm4.2\:\mathrm{km\:s^{-1}}$ (Eri4\_mask3, MJD 59610.1). The difference in the velocity of the binary is inconsistent with zero within the $2\sigma$ uncertainty. Some other member stars have $|\Delta v_{hel}| > 0$, but they are consistent with zero within the $1\sigma$ uncertainty.

In principle, a star could be erroneously identified as a binary if its telluric correction does not follow the systematic trend in $v_{tell}$ as a function of slit position. Systematic trends in $v_{tell}$ are a known artifact in IMACS observations \citep[e.g.,][]{2017ApJ...838....8L, Chiti2022ApJ...939...41C} and likely result from slight rotation of the mask or anamorphic demagnification of the spectrograph. The telluric correction is what accounts for such effects, so a poorly determined telluric correction could lead to an inaccurate heliocentric radial velocity and thus cause the observed variability.

With respect to the binary, we see no evidence that the difference in radial velocity for this star is due to an issue with the telluric correction, as the corrections applied to measurements made using Eri4\_mask1 (MJD 59497.3) and Eri4\_mask1 (MJD 59608.6) are within 3$\sigma$ and 1$\sigma$ of the best fit line respectively. In fact, if we instead use $v_{tell}$ from the fits of $v_{tell}$ versus slit position for both observations, the difference only grows ($\Delta v_{hel} = 8.9\pm3.0\:\mathrm{km\:s^{-1}}$). As such, this object is considered a member but excluded from calculations of systemic velocity and velocity dispersion.

\subsection{Centaurus~I}\label{cen1mems}
\begin{figure*}\centering
\includegraphics[width = 0.9\textwidth]{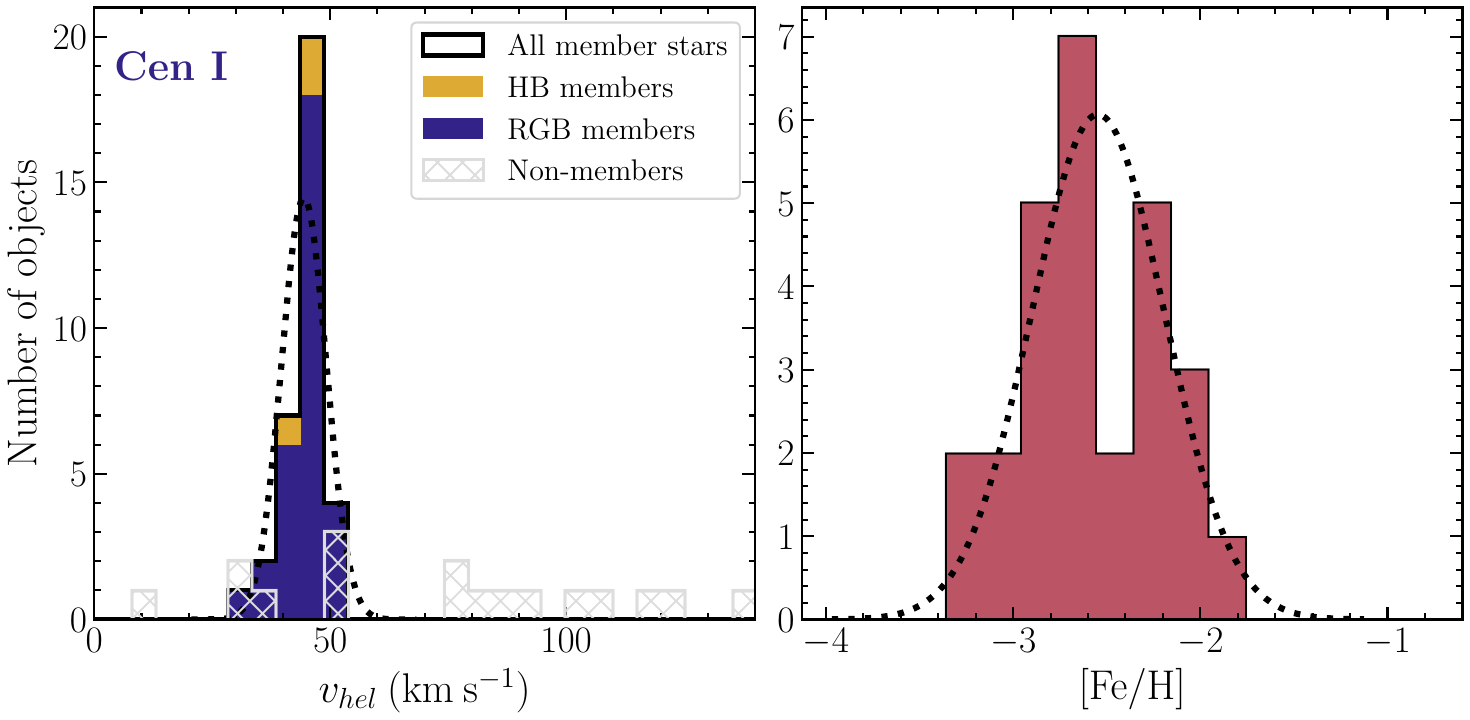}
\caption{(Left) Distribution in velocity of the observed stars near Cen 1 with reliable radial velocity measurements, overplotted with a black dotted line showing the Gaussian model of the systemic velocity and dispersion. The total sample of members is outlined in black, and a stacked histogram of members according to evolutionary stage is plotted in blue (RGB stars) and yellow (HB stars). Non-members are cross-hatched. The range displayed excludes 1 non-member star whose velocity is less than $-50\: \mathrm{km\:s^{-1}}$.
(Right) Distribution of $\mathrm{[Fe/H]}$ of 27 RGB member stars of Cen~I, overplotted with a black dotted line showing the Gaussian model of the systemic metallicity and dispersion. \label{fig:cen1hist} }
\end{figure*}

Observations of Cen~I yielded reliable radial velocity measurements of 62 objects, and of these, we identify 34 member stars. 24 of these member stars have multiple usable spectra. The velocity distribution of the observed stars (Figure \ref{fig:cen1hist}) shows a clear peak at approximately $45 \: \mathrm{km\:s^{-1}}$, which guides the membership selection. Specifically, we consider a star a member of Cen~I if its velocity is $0 \: \mathrm{km\:s^{-1}} < v_{hel} < 60 \: \mathrm{km\:s^{-1}}$, its metallicity is $\mathrm{[Fe/H]} < -1$, its proper motion is $-5 \mathrm{\:mas\:yr^{-1}}< \mu_{\alpha}\cos{\delta} < 5 \mathrm{\:mas\:yr^{-1}}$ and $-3 \mathrm{\:mas\:yr^{-1}}< \mu_{\delta} < 3 \mathrm{\:mas\:yr^{-1}}$, and it lies along the isochrone (a qualitative determination).  Most nonmember stars are excluded on the basis of velocity. Six non-member stars have velocities consistent with Cen~I members, but they are excluded on the basis of proper motion, and one is excluded due to metallicity and its position on the CMD.

The properties of the member stars, as well as observed non-members with reliable radial velocity measurements, are summarized in the bottom row of Figure \ref{fig:summary}. Per the leftmost panel, the member stars (yellow, red, and navy stars) lie within 5 half-light radii of the centroid of Cen~I. They follow the $13\:\mathrm{Gyr}$, $\mathrm{[Fe/H]} = -2.2$ PARSEC isochrone, which is a subpar fit but nonetheless demonstrates coherence along the red giant branch. The rightmost panel shows the grouping of member stars in proper motion. 

\startlongtable
\centerwidetable
\setlength{\tabcolsep}{3pt}
\begin{deluxetable*}{rrrrrrrcrrrrc}
\tabletypesize{\footnotesize}
\tablecaption{Spectroscopically confirmed members of Cen~I.
\label{tab:cen1mems}}
\tablehead{
\colhead{ID} & \colhead{R.A.}  & \colhead{Decl.} & \colhead{$g_0$} & \colhead{$r_0$} & \colhead{$G$}   &\colhead{MJD} & \colhead{Mask} & \colhead{S/N} & \colhead{$v_{hel}$} & \colhead{EW} & \colhead{[Fe/H]}  & \colhead{Notes} \\
           & \colhead{(deg)} &\colhead{(deg)} & \colhead{(mag)} & \colhead{(mag)} & \colhead{(mag)} &            &             &             & \colhead{(km/s)}  & \colhead{(\AA)} & \colhead{(dex)} & }
\startdata
6146256839488355456 & 189.769275 & -40.896825 & 18.75 & 17.98 & 18.05 & 59609.3 & 3 & 38.6  & 49.1$\pm$1.0 & 1.74$\pm$0.61$^a$ & -3.36$\pm$0.14  & RGB      \\
6146246359768228736 & 189.623083 & -40.849067 & 18.66 & 17.79 & 17.87 & 59608.6 & 2 & 65.7  & 44.3$\pm$1.0 & 2.81$\pm$0.29     & -2.86$\pm$0.11  & RGB      \\
...                 & ...        & ...        & ...   & ...   & ...   & 59609.3 & 3 & 41.0  & 46.0$\pm$1.0 & 2.69$\pm$0.31     & -2.91$\pm$0.11  & ...      \\
...                 & ...        & ...        & ...   & ...   & ...   & 58881.3 & 1 & 37.6  & 44.0$\pm$1.0 & 2.86$\pm$0.34     & -2.84$\pm$0.12  & ...      \\
...                 & ...        & ...        & ...   & ...   & ...   & 59410.0 & 2 & 24.4  & 44.2$\pm$1.1 & 2.86$\pm$0.41     & -2.84$\pm$0.14  & ...      \\
6146246359768221440 & 189.618496 & -40.852214 & 19.98 & 19.41 & 19.45 & 59608.6 & 2 & 22.6  & 43.9$\pm$1.1 & 1.91$\pm$0.25     & -3.00$\pm$0.13  & RGB      \\
...                 & ...        & ...        & ...   & ...   & ...   & 58881.3 & 1 & 12.2  & 47.6$\pm$1.2 & 1.76$\pm$0.61     & -3.08$\pm$0.28  & ...      \\
...                 & ...        & ...        & ...   & ...   & ...   & 59410.0 & 2 & 5.0   & 45.2$\pm$2.3 & --                & --              & ...      \\
6146246054826197376 & 189.60315  & -40.857772 & 21.20 & 20.6  & 20.63 & 59608.6 & 2 & 7.6   & 50.8$\pm$1.6 & 1.63$\pm$0.61     & -2.93$\pm$0.31  & RGB      \\
...                 & ...        & ...        & ...   & ...   & ...   & 58881.3 & 1 & 4.6   & 46.9$\pm$2.3 & --                & --              & ...      \\
6146246020466461696 & 189.590371 & -40.857381 & 20.96 & 20.39 & 20.45 & 59608.6 & 2 & 9.7   & 40.0$\pm$1.5 & 2.21$\pm$0.8$^a$  & -2.56$\pm$0.44  & RGB      \\
...                 & ...        & ...        & ...   & ...   & ...   & 58881.3 & 1 & 5.8   & 39.7$\pm$1.8 & --                & --              & ...      \\
6146246016170807552 & 189.59100  & -40.872908 & 20.54 & 19.96 & 19.99 & 59608.6 & 2 & 15.5  & 37.5$\pm$1.2 & 4.51$\pm$0.59     & -1.68$\pm$0.19  & RGB      \\
...                 & ...        & ...        & ...   & ...   & ...   & 59609.3 & 3 & 9.3   & 40.0$\pm$1.5 & 3.92$\pm$0.72     & -1.93$\pm$0.25  & ...      \\
...                 & ...        & ...        & ...   & ...   & ...   & 58881.3 & 1 & 8.4   & 36.7$\pm$1.6 & 4.17$\pm$0.86     & -1.83$\pm$0.30  & ...      \\
6146245917389139712 & 189.579525 & -40.872558 & 20.38 & 19.85 & 19.90 & 59608.6 & 2 & 16.6  & 45.0$\pm$1.2 & 1.88$\pm$0.49     & -2.93$\pm$0.21  & RGB      \\
...                 & ...        & ...        & ...   & ...   & ...   & 59609.3 & 3 & 9.9   & 47.0$\pm$1.3 & --                & --              & ...      \\
...                 & ...        & ...        & ...   & ...   & ...   & 58881.3 & 1 & 9.2   & 45.8$\pm$1.4 & 1.65$\pm$0.31     & -3.07$\pm$0.16  & ...      \\
...                 & ...        & ...        & ...   & ...   & ...   & 59410.0 & 2 & 3.6   & 46.8$\pm$2.9 & --                & --              & ...      \\
6146245878731830912 & 189.567017 & -40.888022 & 20.33 & 19.71 & 19.76 & 59608.6 & 2 & 21.00 & 48.9$\pm$1.1 & 2.67$\pm$0.52     & -2.55$\pm$0.19  & RGB      \\
...                 & ...        & ...        & ...   & ...   & ...   & 58881.3 & 1 & 10.6  & 50.5$\pm$1.3 & 2.61$\pm$0.55     & -2.57$\pm$0.20  & ...      \\
...                 & ...        & ...        & ...   & ...   & ...   & 59410.0 & 2 & 4.1   & 50.2$\pm$2.9 & --                & --              & ...      \\
6146245878731830144 & 189.5685   & -40.894456 & 20.07 & 19.41 & 19.46 & 59608.6 & 2 & 26.3  & 45.4$\pm$1.1 & 2.36$\pm$0.56     & -2.75$\pm$0.21  & RGB      \\
...                 & ...        & ...        & ...   & ...   & ...   & 59609.3 & 3 & 14.9  & 46.0$\pm$1.1 & 2.51$\pm$0.50     & -2.68$\pm$0.18  & ...      \\
...                 & ...        & ...        & ...   & ...   & ...   & 59410.0 & 2 & 4.9   & 45.4$\pm$2.2 & --                & --              & ...      \\
6146245848667777920 & 189.593671 & -40.890114 & --    & --    & 19.78 & 59608.6 & 2 & 19.8  & 45.5$\pm$1.1 & 3.25$\pm$0.47     & -2.27$\pm$0.16  & RGB      \\
...                 & ...        & ...        & ...   & ...   & ...   & 59410.0 & 2 & 3.9   & 45.0$\pm$2.8 & --                & --              & ...      \\
6146245848666932224 & 189.591704 & -40.896467 & 20.33 & 19.67 & 19.75 & 59608.6 & 2 & 20.60 & 44.3$\pm$1.1 & 3.26$\pm$0.46     & -2.28$\pm$0.15  & RGB      \\
...                 & ...        & ...        & ...   & ...   & ...   & 59609.3 & 3 & 11.6  & 45.8$\pm$1.2 & --                & --              & ...      \\
...                 & ...        & ...        & ...   & ...   & ...   & 59410.0 & 2 & 4.1   & 42.9$\pm$3.3 & --                & --              & ...      \\
6146245814309831680 & 189.576479 & -40.897392 & 20.49 & 19.87 & 19.93 & 59609.3 & 3 & 10.3  & 51.6$\pm$1.3 & 2.26$\pm$0.46     & -2.71$\pm$0.18  & RGB      \\
...                 & ...        & ...        & ...   & ...   & ...   & 58881.3 & 1 & 9.6   & 51.1$\pm$1.4 & --                & --              & ...      \\
6146244607421501440 & 189.554492 & -40.894408 & 20.76 & 20.14 & 20.2  & 59608.6 & 2 & 14.4  & 46.3$\pm$1.2 & 3.10$\pm$0.48$^a$ & -2.26$\pm$0.02  & RGB      \\
...                 & ...        & ...        & ...   & ...   & ...   & 59410.0 & 2 & 2.6   & 49.8$\pm$4.6 & --                & --              & ...      \\
6146244405558779392 & 189.556096 & -40.910994 & 20.84 & 20.50 & 20.45 & 59608.6 & 2 & 9.7   & 49.1$\pm$1.6 & 1.93$\pm$0.53     & -2.79$\pm$0.23  & HB       \\
...                 & ...        & ...        & ...   & ...   & ...   & 59609.3 & 3 & 5.4   & 47.9$\pm$2.0 & --                & --              & ...      \\
6146244298187004800 & 189.567346 & -40.927250 & 20.83 & 20.22 & 20.27 & 59608.6 & 2 & 13.5  & 46.0$\pm$1.3 & 2.33$\pm$0.49     & -2.61$\pm$0.19  & RGB      \\
...                 & ...        & ...        & ...   & ...   & ...   & 59609.3 & 3 & 7.9   & 46.6$\pm$1.7 & --                & --              & ...      \\
...                 & ...        & ...        & ...   & ...   & ...   & 58881.3 & 1 & 7.1   & 46.4$\pm$1.7 & --                & --              & ...      \\
6146244195105286144 & 189.54825  & -40.940836 & 20.95 & 20.34 & 20.41 & 59608.6 & 2 & 12.6  & 37.7$\pm$1.3 & 1.92$\pm$0.58     & -2.80$\pm$0.25  & RGB      \\
...                 & ...        & ...        & ...   & ...   & ...   & 59609.3 & 3 & 7.3   & 40.5$\pm$1.6 & --                & --              & ...      \\
...                 & ...        & ...        & ...   & ...   & ...   & 58881.3 & 1 & 6.1   & 42.3$\pm$1.9 & --                & --              & ...      \\
6146244195104572672 & 189.553154 & -40.941889 & 20.37 & 19.91 & 19.92 & 59609.3 & 3 & 10.2  & 48.9$\pm$1.4 & 2.55$\pm$0.54     & -2.58$\pm$0.2   & RGB      \\
6146234235076699392 & 189.633629 & -40.878075 & 20.83 & 20.56 & 20.56 & 59608.6 & 2 & 6.3   & 45.1$\pm$5.0 & --                & --              & HB, RRL  \\
6146234200717222144 & 189.608446 & -40.888353 & 21.32 & 20.79 & 20.79 & 59608.6 & 2 & 7.3   & 41.7$\pm$1.7 & 2.13$\pm$0.41     & -2.60$\pm$0.17  & RGB      \\
6146234162061045632 & 189.627121 & -40.888936 & 18.88 & 17.97 & 18.06 & 59608.6 & 2 & 66.2  & 41.6$\pm$1.0 & 4.62$\pm$0.29     & -2.08$\pm$0.10  & RGB, BIN \\
...                 & ...        & ...        & ...   & ...   & ...   & 58881.3 & 1 & 34.4  & 52.1$\pm$1.0 & 4.68$\pm$0.34     & -2.06$\pm$0.11  & ...      \\
...                 & ...        & ...        & ...   & ...   & ...   & 59410.0 & 2 & 21.8  & 42.8$\pm$1.1 & 4.85$\pm$0.53     & -1.99$\pm$0.16  & ...      \\
6146234127701270400 & 189.62295  & -40.9129   & 20.66 & 20.01 & 20.03 & 59608.6 & 2 & 16.2  & 32.8$\pm$1.2 & 3.78$\pm$0.57     & -1.97$\pm$0.19  & RGB      \\
...                 & ...        & ...        & ...   & ...   & ...   & 58881.3 & 1 & 8.4   & 33.2$\pm$1.5 & --                & --              & ...      \\
6146234093341536640 & 189.604321 & -40.912561 & 18.82 & 17.95 & 18.03 & 59608.6 & 2 & 68.2  & 47.9$\pm$1.0 & 3.46$\pm$0.30     & -2.56$\pm$0.11  & RGB      \\
...                 & ...        & ...        & ...   & ...   & ...   & 59410.0 & 2 & 21.8  & 48.7$\pm$1.1 & 3.79$\pm$0.46     & -2.42$\pm$0.14  & ...      \\
6146234024622071808 & 189.612975 & -40.903703 & 18.68 & 17.75 & 17.83 & 59608.6 & 2 & 77.3  & 44.0$\pm$1.0 & 4.95$\pm$0.27     & -2.01$\pm$0.10  & RGB      \\
...                 & ...        & ...        & ...   & ...   & ...   & 59609.3 & 3 & 44.0  & 45.0$\pm$1.0 & 4.97$\pm$0.31     & -2.00$\pm$0.10  & ...      \\
...                 & ...        & ...        & ...   & ...   & ...   & 58881.3 & 1 & 39.0  & 45.7$\pm$1.0 & 4.99$\pm$0.32     & -1.99$\pm$0.11  & ...      \\
...                 & ...        & ...        & ...   & ...   & ...   & 59410.0 & 2 & 25.5  & 44.5$\pm$1.1 & 4.98$\pm$0.41     & -2.00$\pm$0.12  & ...      \\
6146233857120392960 & 189.637108 & -40.928506 & 21.15 & 20.59 & 20.61 & 59608.6 & 2 & 9.2   & 42.8$\pm$1.5 & 2.19$\pm$0.60     & -2.61$\pm$0.24  & RGB      \\
...                 & ...        & ...        & ...   & ...   & ...   & 59609.3 & 3 & 6.0   & 44.8$\pm$1.7 & --                & --              & ...      \\
6146232547153275136 & 189.564404 & -40.934736 & 18.42 & 17.21 & 17.27 & 59608.6 & 2 & 115.2 & 47.5$\pm$1.0 & 4.81$\pm$0.25     & -2.16$\pm$0.10  & RGB      \\
...                 & ...        & ...        & ...   & ...   & ...   & 59609.3 & 3 & 65.3  & 48.6$\pm$1.0 & 4.68$\pm$0.28     & -2.21$\pm$0.10  & ...      \\
...                 & ...        & ...        & ...   & ...   & ...   & 59410.0 & 2 & 40.0  & 46.5$\pm$1.0 & 4.35$\pm$0.34     & -2.34$\pm$0.11  & ...      \\
6146232272275323392 & 189.592292 & -40.973889 & 20.07 & 19.41 & 19.46 & 59608.6 & 2 & 27.2  & 46.7$\pm$1.1 & 1.61$\pm$0.33     & -3.16$\pm$0.17  & RGB      \\
...                 & ...        & ...        & ...   & ...   & ...   & 58881.3 & 1 & 13.5  & 47.9$\pm$1.2 & 1.46$\pm$0.28     & -3.27$\pm$0.17  & ...      \\
6146232100476627456 & 189.590983 & -40.978708 & 20.89 & 20.27 & 20.31 & 59608.6 & 2 & 13.8  & 52.2$\pm$1.3 & 2.98$\pm$0.59     & -2.28$\pm$0.21  & RGB      \\
6146230558584952192 & 189.606167 & -41.082858 & 21.0  & 20.76 & 20.67 & 59608.6 & 2 & 6.4   & 42.5$\pm$2.2 & --                & --              & HB       \\
6146225950085206912 & 189.477496 & -41.040197 & 19.28 & 18.52 & 18.60 & 59608.6 & 2 & 50.0  & 47.2$\pm$1.0 & 2.37$\pm$0.22     & -2.92$\pm$0.11  & RGB      \\
...                 & ...        & ...        & ...   & ...   & ...   & 59410.0 & 2 & 11.6  & 44.8$\pm$1.3 & 2.30$\pm$0.76     & -2.95$\pm$0.29  & ...      \\
11024300316381      & 189.608417 & -40.886558 & 21.88 & 21.40 & --    & 59608.6 & 2 & 4.2   & 44.0$\pm$2.8 & --                & --              & RGB      \\
11024300262356      & 189.629417 & -40.834544 & 21.33 & 20.78 & --    & 59608.6 & 2 & 5.9   & 42.3$\pm$2.3 & --                & --              & RGB      \\
11024300234422      & 189.590458 & -40.865800 & 21.48 & 20.98 & --    & 59608.6 & 2 & 5.7   & 43.5$\pm$2.1 & --                & --              & RGB      \\
11024300234367      & 189.544004 & -40.960144 & 21.92 & 21.39 & --    & 59608.6 & 2 & 5.4   & 48.2$\pm$2.3 & --                & --              & RGB      \\
11024300030965      & 189.531242 & -40.913783 & 21.70 & 21.14 & --    & 59608.6 & 2 & 5.7   & 36.4$\pm$2.1 & --                & --              & RGB      \\
...                 & ...        & ...        & ...   & ...   & ...   & 59609.3 & 3 & 3.3   & 35.5$\pm$2.9 & --                & --              & ...      
\enddata
\tablenotetext{a}{The EW of one of the three CaT II lines of this star is derived using the EW relationships described in Section \ref{sec:met}, as it cannot be measured due to a chip gap or fit issue affecting one line. The other two lines are measured with \texttt{mpfit}.}
\end{deluxetable*}

Amongst the 34 member stars, three HB stars (yellow stars in Figure \ref{fig:summary}) are identified according to DELVE photometry (center left panel, bottom row of Figure \ref{fig:summary}). The five faintest member stars of Cen~I are present only in DELVE photometry (outlined in magenta in Figure \ref{fig:summary}), and one star is present only in \textit{Gaia} photometry (outlined in cyan). We also measure the metallicity of 27 of the 31 RGB stars, presented in the right panel of Figure \ref{fig:cen1hist} and marked as red stars in Figure \ref{fig:summary}.

One RR Lyrae, identified in \citet{2021AJ....162..253M}, is present in the 34-star sample (outlined in Figure \ref{fig:summary} in green). We exclude this star from any calculations involving radial velocity. No additional stars in our sample were identified as variable using the \textit{Gaia} DR3 variability catalogue \citep{2022arXiv220606278C}. 

Comparison of the velocities measured in different epochs suggests one possible binary. This object (flagged in Table \ref{tab:cen1mems} as BIN) was observed in February 2020 using Cen1\_mask1 (MJD 58881.3), July 2021 using Cen1\_mask2 (MJD 59410.0), and January 2022 using Cen1\_mask2 (MJD 59608.6). The difference in velocity between February 2020 (Cen1\_mask1) and January 2022 (Cen1\_mask2) is $|\Delta v_{hel}| = 9.3\pm1.5\:\mathrm{km\:s^{-1}}$, which is inconsistent with zero at $>3\sigma$. There is no appreciable change in velocity between July 2021 and January 2022. Given its varying velocity, we exclude this star from the calculation of the velocity dispersion. Some other member stars have $|\Delta v_{hel}| > 0$, but all are consistent with zero within the 3$\sigma$ uncertainty and most are consistent within $1\sigma$.

There is also evidence of a systematic offset of $\approx 1\:\mathrm{km\:s^{-1}}$ between Cen1\_mask3 (MJD 59609.3) and the other measurements. The median offset between velocity measurements made using Cen1\_mask3 (MJD 59609.3) and Cen1\_mask1 (MJD 58881.3), Cen1\_mask2 (MJD 59410.0), and Cen1\_mask2 (MJD 59608.6) is $0.5\pm0.7$, $1.2\pm0.5$, and $1.3\pm0.4\:\mathrm{km\:s^{-1}}$, determined with the 7, 6, and 12 stars with radial velocity measurements in common with Cen1\_mask3. In comparison, the offset between measurements made in the same observing run as Cen1\_mask3 using a different mask, Cen1\_mask2 (MJD 59608.6), and measurements from a different run using Cen1\_mask2 (MJD 59410.0) is $0.5\pm0.5\:\mathrm{km\:s^{-1}}$. A median offset of $>1\:\mathrm{km\:s^{-1}}$ between observations of the same objects on the same nights using different masks is unusual and suggests an unaccounted-for systematic error. Given that the offset exists between measurements made using Cen1\_mask3 and measurements made with different masks during both the same run and a different run, we choose to apply a $-1\:\mathrm{km\:s^{-1}}$ correction to Cen1\_mask3 measurements. All reported and tabulated values include this correction. Although we apply this correction, no statistically significant differences exist for any reported values between calculations with this correction and calculations without. It also has no effect on membership determination or identification of binaries.

\section{Systemic parameters}\label{sec:sys}
\subsection{Systemic Velocity and Velocity Dispersion}\label{sec:vdisp}

To calculate the velocity dispersion and systemic velocity of the sample, we again employ \texttt{emcee}. Assuming the velocities of each star are normally distributed about a mean systemic velocity $v_{sys}$ with intrinsic dispersion $\sigma_{v}$, we can adopt a two-parameter Gaussian likelihood function as in \citet{2006AJ....131.2114W}, which considers both intrinsic dispersion and measurement uncertainty:
\begin{equation}
    \log{\mathcal{L}}\propto -\frac{1}{2} [\sum_{n = 1}^{N} \log{(\sigma_{v}^2 + \sigma_{v_i}^2)} + \sum_{n = 1}^{N} \log{\frac{(v_i - v_{sys})^2}{\sigma_{v}^2 + \sigma_{v_i}^2}}]
    \label{eqn:disp}
\end{equation}

\noindent where $v_{sys}$ denotes the systemic heliocentric velocity, $\sigma_{v}$ denotes the velocity dispersion, and $v_i$ refers to an individual star with velocity $v_i$ and uncertainty $\sigma_{v_i}$. We use 50 walkers, each taking 2000 steps, with a burn-in period of 100 steps, and we use a Jeffreys prior for the velocity between the minimum and maximum heliocentric velocity of the sample and a log-uniform prior with bounds $-2 < \log{\sigma_{v}} < 2$ for the dispersion. We exclude the known RR Lyrae and suspected binary in each system from these calculations, as their velocities vary with time and could artificially change the dispersion. We report the median value of the posterior and calculate uncertainty from the standard deviation.

With this method, Eri~IV has a measured systemic velocity of $v_{sys} = -31.5^{+1.3}_{-1.2}\:\mathrm{km\:s^{-1}}$ and velocity dispersion $\sigma_{v} = 6.1^{+1.2}_{-0.9}\:\mathrm{km\:s^{-1}}$. Cen~I has a measured systemic velocity of $44.9^{+0.8}_{-0.8}\:\mathrm{km\:s^{-1}}$ and a velocity dispersion of $4.2^{+0.6}_{-0.5} \:\mathrm{km\:s^{-1}}$.

To test the robustness of the velocity dispersion and systemic velocity, we performed a jackknife test on both systems. In a jackknife test, one sample at a time is removed from the population, and the parameters are recalculated from the resulting subsample. We use it to evaluate the sensitivity of the systemic parameters to individual stars. For both Eri~IV and Cen~I, the systemic velocity and velocity dispersion do not show any particular influence from one star; the values of these parameters in all jackknife subsamples fall within the reported uncertainty of $v_{sys}$ and $\sigma_{v}$ of the full sample (the 16th/84th percentiles of the distribution). This indicates that the systemic velocity and velocity dispersion are unlikely to be strongly affected by contamination by a field star or undetected binary.

For both systems, we recalculated $v_{sys}$ and $\sigma_v$ including the RR Lyrae and suspected binary, in the event that they are misidentified. The resulting velocities and velocity dispersions, tabulated in Table \ref{tab:subsamp}, are consistent with the parameters of the fiducial sample that excludes them for both Eri~IV and Cen~I.

For Eri~IV, we also consider the effect of the HB stars on the systemic properties, given the somewhat large velocity dispersion and the fact that the two BHBs have the highest velocities of the entire sample. The velocity dispersion drops from $6.1_{-0.9}^{+1.2}\:\mathrm{km\:s^{-1}}$ to $5.7_{-0.8}^{+1.0}\:\mathrm{km\:s^{-1}}$ when the HB stars are excluded from the sample (this result is tabulated in Table \ref{tab:subsamp}). In general, the S/N of spectra of the BHBs is low ($\approx 6$) and the fit uncertainty is large, so their contribution to the calculation of systemic properties is small. The third HB star has a high S/N, but its velocity is consistent with the systemic velocity, so its contribution is also small. For Cen~I, the velocities of the two non-variable HB stars are within $1\sigma$ of the mean velocity, so while this test was performed (and its results tabulated in Table \ref{tab:subsamp}), the effect is negligible.

We also repeated this analysis on only non-variable member stars with measured metallicities in each galaxy, which have the most confident membership determinations. The velocity and velocity dispersion of these subsamples (Table \ref{tab:subsamp}) are consistent with the parameters of the fiducial samples for Eri~IV and Cen~I. The consistency of the systemic parameters for these various subsamples of member stars indicates that they are relatively robustly determined, especially with respect to contamination.

\begin{figure}\centering
\includegraphics[width = \columnwidth]{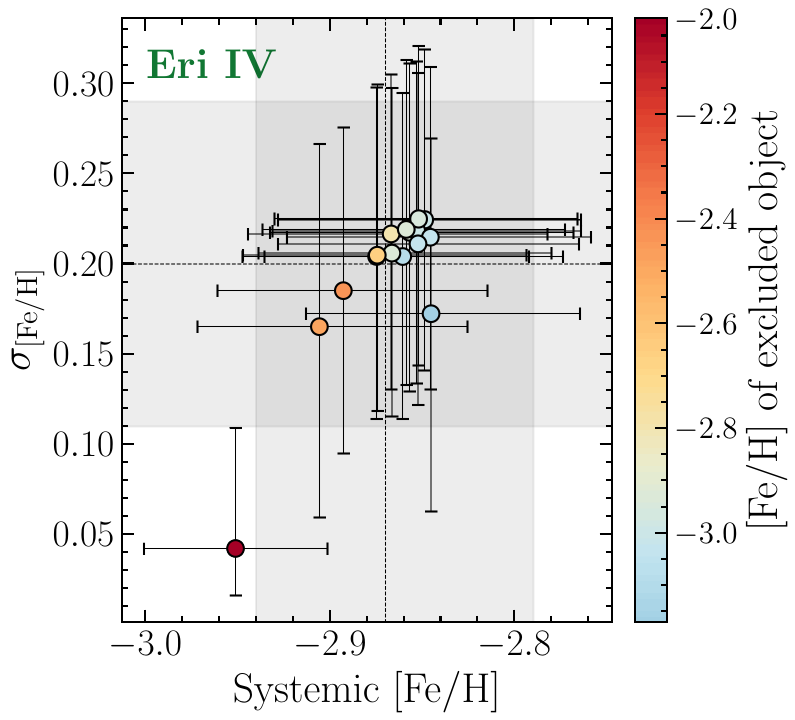}
\caption{Jackknife test for metallicity parameters of Eri IV. Highlighted in gray are the 16th/84th percentile of the full sample. Each point represents the systemic parameters of a subsample that excludes one star. The metallicity of that excluded star is represented by the color of the point. The colorbar is centered at the systemic value measured from the full sample. Excluding the most metal-rich member star results in metallicity and metallicity dispersion discrepant with the full sample, illustrated by the darkest red point in the bottom left.\label{fig:eri4jack}}
\end{figure}

\subsection{Systemic Metallicity and Metallicity Dispersion}\label{sec:mdisp}
As with the systemic velocity and velocity dispersion, \texttt{emcee} is used to calculate the mean metallicity and metallicity dispersion of the 16-star sample of Eri~IV and the 27-star sample of Cen~I. We use 50 walkers, each taking 2000 steps, with a burn-in of 100 steps. The initial position is determined by a maximum likelihood estimate, so longer sampling or burn-in is not necessary. The likelihood is the same as in Section \ref{sec:vdisp}, adapted for metallicity and metallicity dispersion. We find the metallicity of Eri~IV to be $\mathrm{[Fe/H]} = -2.87^{+0.08}_{-0.07}$ and the dispersion to be $\sigma_{\mathrm{[Fe/H]}} = 0.20\pm0.09$. For Cen~I, we find a metallicity of $\mathrm{[Fe/H]} = -2.57\pm0.08$ and a dispersion of $\sigma_{\mathrm{[Fe/H]}} = 0.38^{+0.09}_{-0.07}$. The possible non-Gaussianity of the Eri~IV metallicity distribution function is addressed in Section \ref{sec:mdf}, but we assume a normal metallicity distribution in this analysis, a choice consistent with the literature \citep[e.g.,][]{Koposov2015:2015ApJ...811...62K,2015ApJ...808...95S,Lai2011:2011ApJ...738...51L,2017ApJ...838...11S,Chiti2022ApJ...939...41C}, to facilitate comparison with other dwarf galaxies.

\begin{figure*}\centering
\includegraphics[width = 0.9\textwidth]{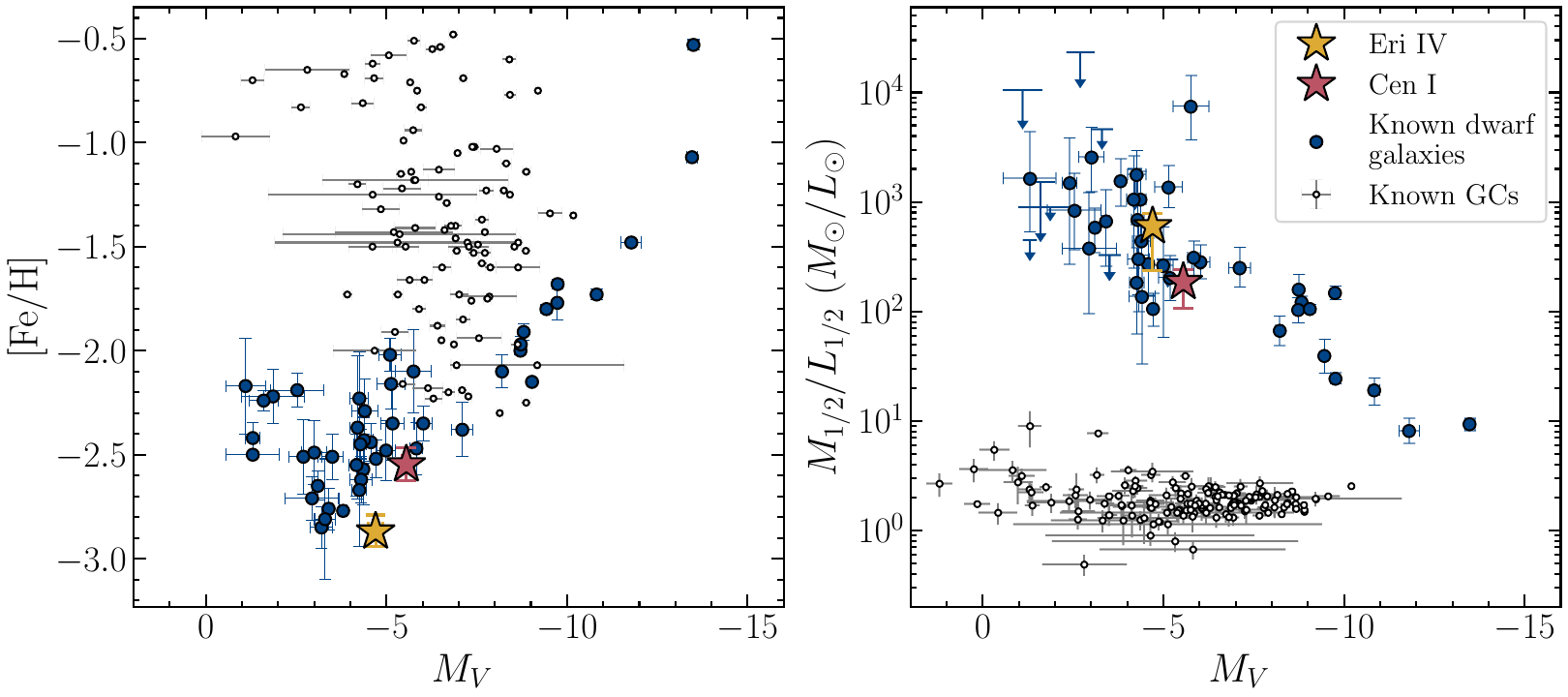}\hfil

\caption{(Left) Absolute V-band magnitude ($M_V$) vs. metallicity ($\mathrm{[Fe/H]}$) for the known dwarf galaxies (\protect \citealt{2022ApJ...940..136P}, except for Boo~II and Aqu~II, which use updated values from \protect \citealt{2023arXiv230203708B}) and globular clusters \protect \citep{2018MNRAS.478.1520B, 2020PASA...37...46B} with a measured metallicity. Eri~IV is more metal-poor than other dwarfs at similar absolute magnitude; Cen~I is consistent.
(Right) Absolute V-band magnitude vs. mass-to-light ratio ($M/L$). Eri~IV and Cen~I are both consistent with the population of dwarf galaxies. 
\label{fig:mvcomparison}}
\end{figure*}

The metallicity and metallicity dispersion of Cen~I are typical for a dwarf galaxy at a magnitude of $M_V\sim-5.55$. The spread in metallicity is reflected in the widening of the RGB on the color-magnitude diagram (center bottom of Figure \ref{fig:summary}). The spread in metallicity with color is most clear in the five bright RGB stars at $g\sim18.75$. The bluest star has a metallicity of $-3.36\pm0.14$ and the reddest has a metallicity of $-2.06\pm0.07$.

Compared to other known dwarf galaxies with measured metallicities, Eri~IV is more metal-poor (Figure \ref{fig:mvcomparison}). Its metallicity dispersion is smaller than most dwarfs at similar magnitudes, but at faint magnitudes, the scatter increases, and Eri~IV falls within this general trend.

The most confidently determined members are the non-variable members with measured metallicities (15 stars in Eri~IV and 26 in Cen~I). When calculating the metallicity and metallicity dispersion with these subsamples, we find no statistically significant difference from the systemic parameters calculated with the fiducial sample (see Table \ref{tab:subsamp}), although the metallicity dispersion of Eri~IV is less significantly resolved with this sample (consistent with zero within $2\sigma$).

We also performed a jackknife test to evaluate the robustness of the mean metallicity and metallicity dispersion. Cen~I displays no particular influence from any one star. Looking at the extremes, the most metal-rich star in our Cen~I sample has $\mathrm{[Fe/H]} = -1.79\pm0.14$. Predictably, its exclusion lowers the dispersion and lowers the metallicity, but the effect is not statistically significant. The same is true of the most metal-poor star.

With respect to Eri~IV, the most metal-rich star in the sample has an appreciable effect on its mean metallicity and dispersion, illustrated in Figure \ref{fig:eri4jack}. This star has a metallicity of $-2.01\pm0.22$, which is $2.77\sigma$ from the systemic metallicity. By removing it, the dispersion drops to $0.04^{+0.07}_{-0.03}$ and the metallicity to $\mathrm{[Fe/H]} = -2.95\pm0.05$, which is discrepant with the parameters of the full sample. Upon repeat inspection of this object's fits and diagnostics, we found no credible reason to exclude it as a member. The quality of the fit is high (S/N$ = 13.08$), it lies within 1 half-light radius of the centroid, it is consistent with other members on the CMD, and its proper motion and radial velocity are consistent with the systemic properties within $1\sigma$ and $2\sigma$ respectively. There is also no evidence of a strong Mg line at 8807 \AA, which can be used to differentiate between giant and dwarf stars due to its sensitivity to surface gravity \citep{2012A&A...539A.123B}. Additionally, although the metallicity of this star is high in comparison to members of Eri~IV, stars at this metallicity are not uncommon in UFDs. Approximately 19\% of the available literature sample of stars in UFDs have metallicity $>-2.0$ \citep{Fu2023:2023arXiv230606260F}. Given these factors, we think this star is more likely to be a member star of Eri~IV than a field star.

\begin{deluxetable*}{lllllll}
\tablewidth{0pt}
\tablecaption{Comparison of systemic properties for different subsamples of Eri~IV and Cen~I.\label{tab:subsamp}}
\tablehead{
\colhead{Sample Name}                 & \colhead{Sample Makeup}                           & \colhead{$v_{sys}$ }                & \colhead{$\sigma_{v}$ }       & \colhead{[Fe/H]}                    & \colhead{$\sigma_{\mathrm{[Fe/H]}}$} \\
 & \colhead{} &  \colhead{$\mathrm{km\:s^{-1}}$}     & \colhead{$\mathrm{km\:s^{-1}}$}     & \colhead{dex }                      & \colhead{dex }
}
\startdata
\multicolumn{6}{c}{Eri~IV}\\ 
\midrule
Full $v_{sys}$   (fiducial)   & 23 RGB + 3 HB                    & $-31.5_{-1.2}^{+1.3}$  & $6.1_{-0.9}^{+1.2}$    & ...                       & ... \\
Full metallicity (fiducial)   & 16 RGB                           & ...                       & ...                       & $-2.87^{+0.08}_{-0.07}$   & $0.20\pm0.09$ \\
RGB only	                 & 23 RGB                           & $-32.3\pm1.3$  & $5.7_{-0.8}^{+1.0}$    & ...                       & ... \\
Without most metal-rich member star & 22 RGB + 3 HB, 15 RGB            & $-32.0\pm1.2$   & $5.6^{+1.2}_{-0.9}$    & $-2.95\pm0.05$            & $0.04^{+0.07}_{-0.03}$ \\ 
Non-variable members with [Fe/H] measurement  & 15 RGB           & $-32.9\pm1.5$              & $5.5^{+1.3}_{-0.9}$       & $-2.91^{+0.08}_{-0.06}$   & $0.17\pm0.10$ \\
\midrule
\multicolumn{6}{c}{Cen I}\\
\midrule
Full $v_{sys}$   (fiducial)   & 31 RGB + 1 HB    & $44.9\pm0.8$            & $4.2^{+0.5}_{-0.6}$       & ...                       & ... \\
Full metallicity (fiducial)   & 27 RGB           & ...                       & ...                       & $-2.57\pm0.08$   & $0.38^{+0.07}_{-0.05}$ \\
RGB only	                 & 31 RGB           & $44.8\pm0.8$             & $4.3^{+0.7}_{-0.6}$    & ...                       & ... \\
Non-variable members with [Fe/H] measurement  & 26 RGB           & $45.3\pm0.8$              & $4.2^{+0.7}_{-0.6}$       & $-2.59\pm0.08$   & $0.37^{+0.07}_{-0.05}$ \\
\bottomrule
\enddata
\end{deluxetable*}

At the other extreme, we examined the most metal-poor star found in Eri~IV, which has a metallicity of $-3.18\pm0.12$ dex. This is the brightest star in our entire sample, and it is only found in \textit{Gaia} photometry, so its metallicity is calculated with a different photometric system than the rest. It is highly unlikely, however, that the photometric system influenced the low metallicity, as the difference in metallicities calculated using \textit{Gaia} versus DELVE photometry is negligible (see Section \ref{sec:mdisp}). Additionally, this star is among nine other stars with $\mathrm{[Fe/H]} < -2.9$, measured from spectra with a median S/N of 21, so its individual effect on the systemic properties is small. 

\subsection{Systemic Proper Motion}\label{sec:pmsys}
The systemic proper motion 
($\mu_{\alpha}\cos{\delta}$ and $\mu_{\delta}$) is measured similarly to the other systemic properties. We select member stars with astrometric solutions in \textit{Gaia} EDR3 \citep{Lindegren2021:2021A&A...649A...2L} that satisfy \texttt{ruwe}$< 1.4$ and \texttt{astrometric\_excess\_noise}$<$2 (21 in Eri~IV and 22 in Cen~I). The likelihood is a multivariate normal distribution, and we include correlation between $\mu_{\alpha}\cos{\delta}$ and $\mu_{\delta}$ as well as uncertainties. The priors for $\mu_{\alpha}\cos{\delta}$ and $\mu_{\delta}$ are uniform between the minimum and maximum values of the observed values of the parameter. We use 20 walkers taking 2000 steps after a burn-in of 500 steps. 

With this method, Eri~IV has a systemic proper motion of $(\mu_{\alpha}\cos{\delta},\mu_{\delta}) = (0.22\pm0.06, -0.11\pm0.05)\mathrm{\:mas\:yr^{-1}}$, which is in good agreement with the values found in the discovery paper \citep{2021ApJ...920L..44C}, which uses stars with a high probability of membership based on photometry and astrometry as in \citet{2022ApJ...940..136P}. For Cen~I, we find a systemic proper motion of $(\mu_{\alpha}\cos{\delta},\mu_{\delta}) = (-0.14\pm0.05, -0.19\pm0.04)\mathrm{\:mas\:yr^{-1}}$. These values agree well with values found by \citet{2022ApJ...940..136P}, \citet{2022A&A...657A..54B}, and \citet{2020RNAAS...4..229M}, which also use \textit{Gaia} EDR3, but not those originally reported by \citet{2020ApJ...890..136M}, which uses \textit{Gaia} DR2. The discrepancy (which is at approximately the $2\sigma$ level) is due to the improved precision of proper motion measurements and reduced systematics in EDR3 compared to DR2.

\section{Discussion}\label{sec:discussion}
\subsection{Mass-to-Light Ratio}\label{sec:mass2light}
The division between globular clusters and dwarf galaxies is often drawn most clearly by dark matter content: globular clusters are thought to be dominated by baryonic matter and dwarf galaxies by dark matter \citep{2012AJ....144...76W}. Eri~IV and Cen~I were both identified as dwarf galaxy candidates based on their physical size, and we use the measured velocity dispersion to calculate the mass-to-light ratio within the half-light radius to confirm whether they are dark matter-dominated. We first estimate the enclosed mass per \citet{2010MNRAS.406.1220W}:

\begin{equation}
    M_{1/2} \approx 930\times\frac{\sigma_{v}^2}{\mathrm{km\:s^{-1}}}\times \frac{r_{1/2}}{\mathrm{pc}}\times M_{\odot}
\end{equation}

 Using the half-light radius and luminosity from \citet{2021ApJ...920L..44C}, we find the enclosed mass of Eri~IV  to be $M_{1/2} = 1.90^{+1.05}_{-0.70}\times10^{6}\:M_{\odot}$ and the mass-to-light ratio to be $M_{1/2}/L_{1/2} = 598^{+361}_{-228} M_{\odot}/L_{\odot}$. Repeating this for Cen~I using the half-light radius and luminosity from \citet{2020ApJ...890..136M}, we measure an enclosed mass of $1.36^{+0.39}_{-0.28}\times10^{6}\:M_{\odot}$ and a mass-to-light ratio $M_{1/2}/L_{1/2} = 140^{+60}_{-42} M_{\odot}/L_{\odot}$. 
 
 The stratification of globular clusters and dwarf galaxies can be seen in Figure \ref{fig:mvcomparison}. Eri~IV and Cen~I both exhibit high mass-to-light ratios typical of dwarf galaxies at similar magnitudes, which are much greater than the mass-to-light ratio of $<~3 M_{\odot}/L_{\odot}$ characteristic of MW globular clusters \citep{2020PASA...37...46B}. We therefore conclude that both Eri~IV and Cen~I are dark matter-dominated dwarf galaxies.

\subsection{Orbit}\label{sec:orbit}
We can also examine orbital properties to further characterize and understand Eri~IV and Cen~I. Using the systemic velocity and the proper motion of the member stars, we determined the orbital parameters of Eri~IV and Cen~I using \texttt{galpy} \citep{2015ApJS..216...29B}. We add the Large Magellanic Cloud (LMC) and its effect on the barycentric acceleration of the Milky Way to the default \texttt{galpy} potential, MWPotential2014, which includes a bulge model, a Miyamoto Nagai disk, and a Navarro-Frenk-White dark matter halo \citep{2015ApJS..216...29B}. We approximate the effect of the LMC by first modelling it as a Hernquist potential with a mass of $10^{11}\:M_{\odot}$ \citep{2019MNRAS.487.2685E} and computing its orbit given a non-accelerating Milky Way and in the presence of dynamical friction. We then compute the acceleration of the Milky Way due to this model. We adopt as the solar motion $(U, V, W)_{\odot} = (11.1^{+0.69}_{-0.75}, 12.24^{+0.47}_{-0.47}, 7.25^{+0.37}_{-0.36})\:\mathrm{km\:s^{-1}}$ \citep{2010MNRAS.403.1829S}, circular velocity $v_o = 220\:\mathrm{km\:s^{-1}}$, and distance to the Galactic center $R_o = 8\:\mathrm{kpc}$ \citep{2012ApJ...759..131B}.

\begin{figure*}\centering
\includegraphics[width=0.9\textwidth]{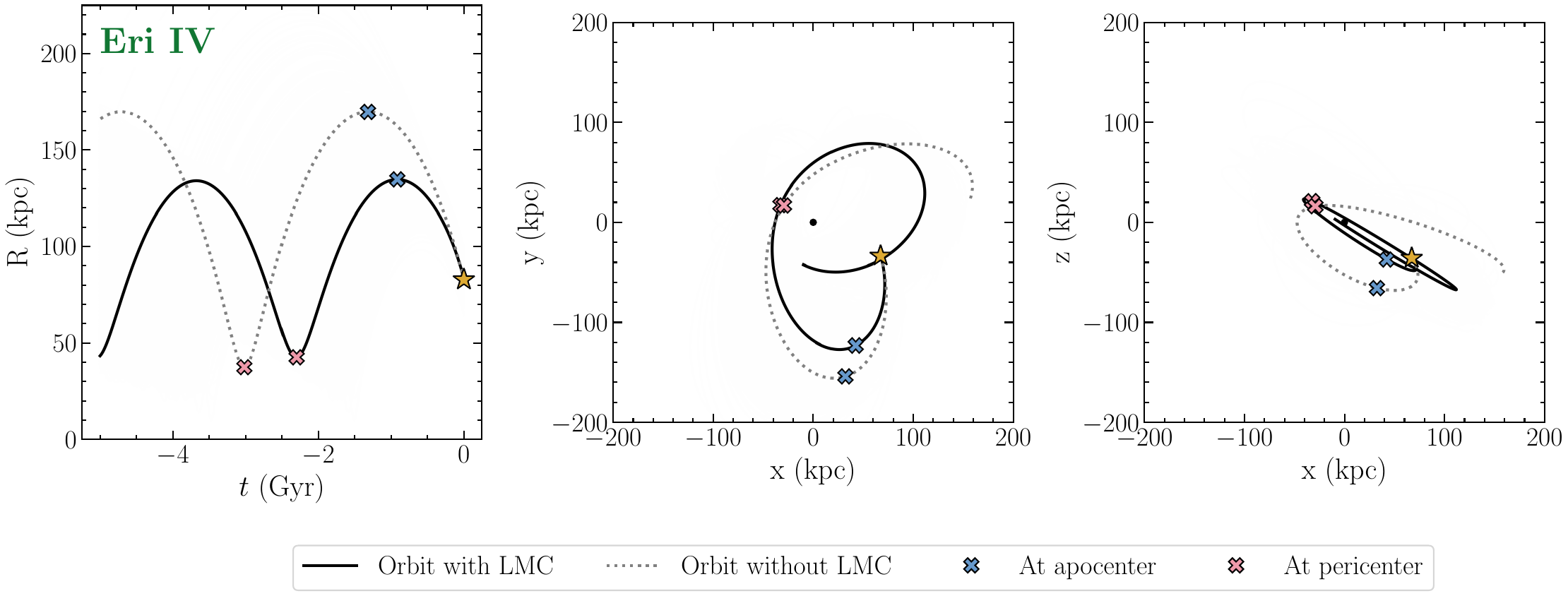}
\caption{(Left) Radial distance from the Galactic center over the past 5 Gyr. Eri~IV is evidently moving towards its pericenter. The opaque black line is the fiducial orbit and the transparent lines are an ensemble of 1000 orbit instances. The dotted line shows the orbit using the default potential that does not include the effect of the LMC. (Center, Right) Projection of Eri~IV's orbit over the past 5 Gyr. 
\label{fig:eri4orb}}
\end{figure*}

\begin{figure*}\centering
\includegraphics[width=0.9\textwidth]{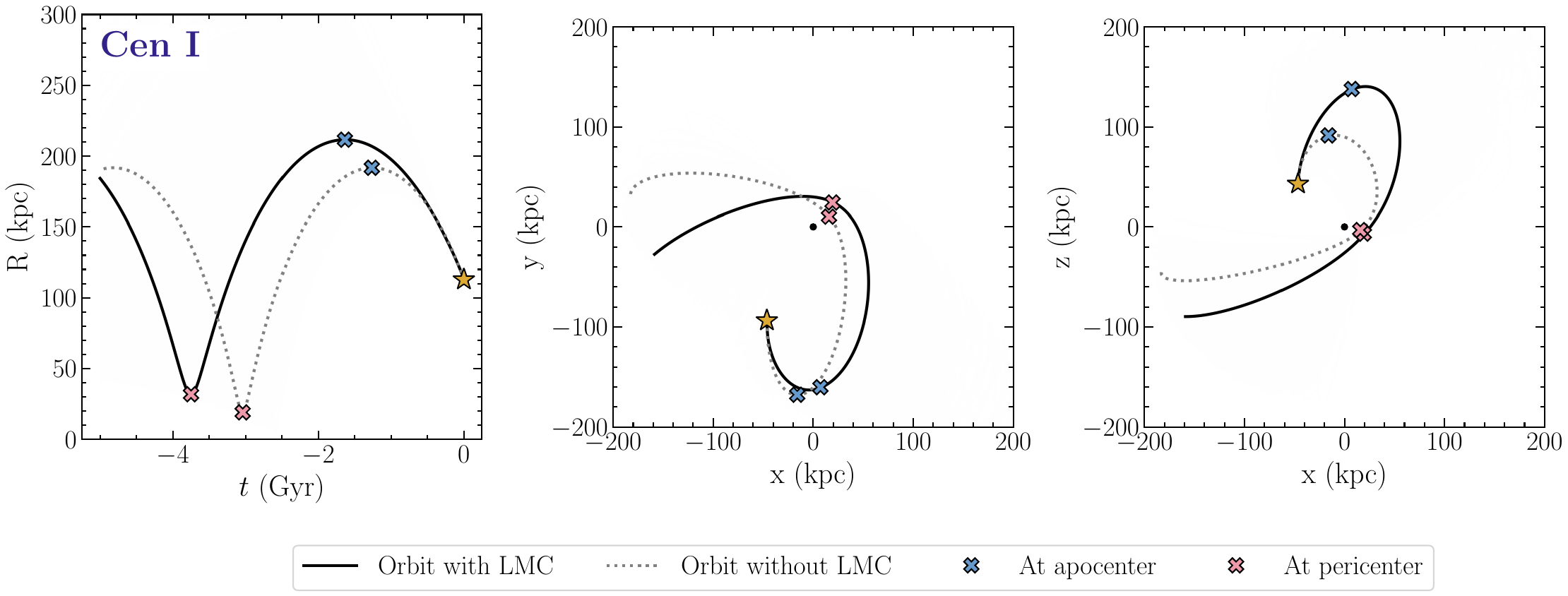}
\caption{(Left) Radial distance from the Galactic center over the past 5 Gyr. The opaque black line is the fiducial orbit and the transparent lines are an ensemble of 1000 orbit instances. The dotted line shows the orbit using the default potential that does not include the effect of the LMC. (Center, Right) Projection of Cen~I's orbit over the past 7.5 Gyr. 
\label{fig:cen1orb}}
\end{figure*}

For each system, we instantiated 1000 orbits with initial conditions $(\mathrm{R.A},\:\mathrm{Decl.},\:D,\:\mu_{\alpha^*},\:\mu_{\delta},\: v_{sys})$ randomly sampled from their probability distributions, which are approximated by a Gaussian distribution. For parameters with asymmetric uncertainty, we sample first from a split-normal distribution and then approximate this as a Gaussian. The reported uncertainty of the orbital parameters is the 16th/84th percentile of the distribution of the 1000 orbit instances. A projection of these orbit instances in $x-y$ and $x-z$ and the period are presented in Figure \ref{fig:eri4orb} for Eri~IV and Figure \ref{fig:cen1orb} for Cen~I, with the most recent apocenter and pericenter marked. 

In integrating back 5 Gyr with this potential, we find that Eri~IV was at its apocenter of $r_{apo} = 135^{+24}_{-13}\mathrm{\:kpc}$ approximately $~1\mathrm{\:Gyr}$ ago, and at its pericenter of $r_{peri} = 43\pm11\mathrm{\:kpc}$ approximately $~2.3\mathrm{\:Gyr}$ ago. It has an orbital eccentricity of $0.52^{+0.07}_{-0.03}$.

We find that Cen~I has an orbital eccentricity of $0.74^{+0.05}_{-0.04}$. It reached its apocenter $r_{apo} = 212^{+39}_{-17}\mathrm{\:kpc}$ approximately $~1.6\mathrm{\:Gyr}$ ago and its pericenter $r_{peri} = 32^{+12}_{-8}\mathrm{\:kpc}$ approximately $~3.3\mathrm{\:Gyr}$ ago. Cen~I is evidently moving towards its pericenter.

We repeat this analysis using MWPotential2014 (a model without the barycentric acceleration of the MW due to the infall of the LMC) and find no statistically significant change in pericenter for either system. There is also little change in apocenter for Cen~I (Figure \ref{fig:cen1orb}). The effect on Eri~IV is somewhat more significant; the apocenter and eccentricity both increase by $\sim20\%$ using MWPotential2014, making them marginally inconsistent with parameters from the fiducial potential model (Figure \ref{fig:eri4orb}). We also repeat this analysis using the potential from \citet{2017MNRAS.465...76M}, which is a heavier model for the MW that assumes a somewhat different circular velocity ($v_o = 232.8\:\mathrm{km\:s^{-1}}$) and distance to the Galactic center ($R_o = 8.20\:\mathrm{kpc}$) \citep{2012ApJ...759..131B}. We also include the barycentric acceleration of the MW due to the infall of the LMC in the model. With this model, there are no significant changes in pericenter, apocenter, or eccentricity. The pericenter distance, which is relevant to questions of tidal distruption, does decrease for both systems (Eri~IV: $r_{peri} = 39^{+9}_{-8}\mathrm{\:kpc}$; Cen~I: $r_{peri}=27^{+10}_{-7}\mathrm{\:kpc}$), but this effect is not statistically meaningful. 

\begin{figure}
\includegraphics[width=\columnwidth]{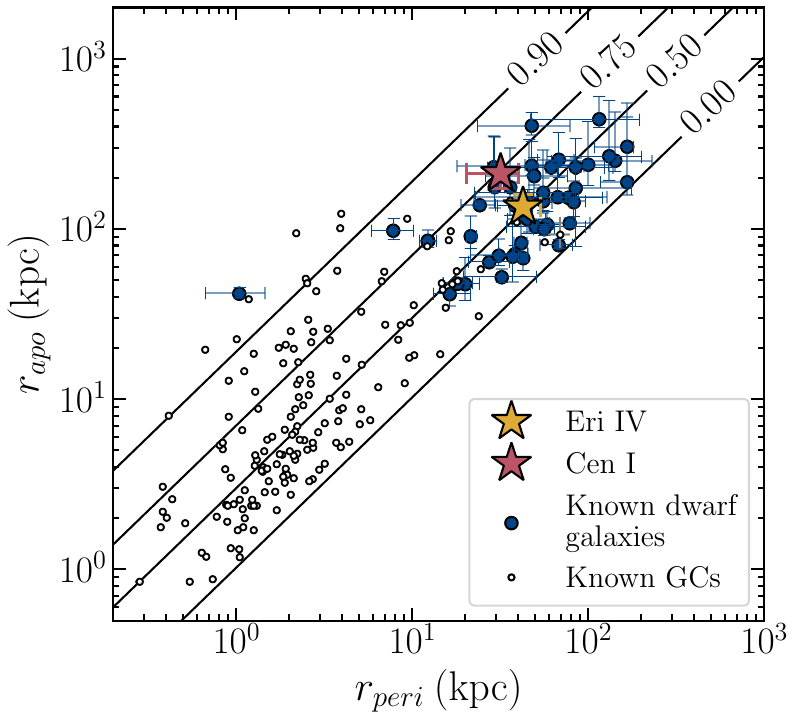}
\caption{Pericenter distance ($r_{peri}$) vs. apocenter distance ($r_{apo}$) for a population of known Milky Way dwarf galaxies (blue circles) from \protect \citet{2022ApJ...940..136P}, except for Boo~II and Aqu~II, which use updated values from \protect \citet{2023arXiv230203708B}, and globular clusters (small open circles) from the Galactic Globular Cluster Database \protect \citep{2018MNRAS.478.1520B, 2020PASA...37...46B}. Lines of constant eccentricity are marked. Eri~IV (yellow star) and Cen~I (pink star) are consistent with the population of dwarf galaxies. \label{fig:apoperiecc}}
\end{figure}

Overall, the orbital parameters of Cen~I under a variety of potential models are consistent within $1\sigma$. The pericenter of Eri~IV is consistent within $1\sigma$ under different potential models, and the apocenter and eccentricity are consistent within $2\sigma$. In particular, the orbital parameters of these systems are largely unaffected by including/excluding the effect of the LMC on the barycentric acceleration of the Milky Way in the potential model. Furthermore, their orbital parameters are similar to other satellites (Figure \ref{fig:apoperiecc}).

\subsection{Tidal Disruption}\label{sec:tidal}
\begin{figure*}\centering
\includegraphics[width = \columnwidth]{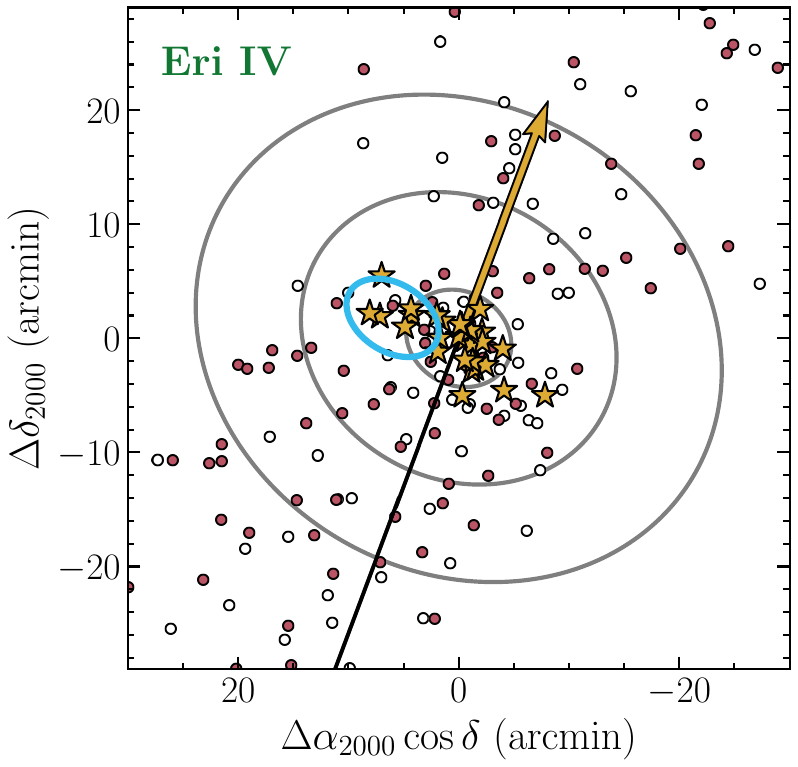}\hfil
\includegraphics[width = \columnwidth]{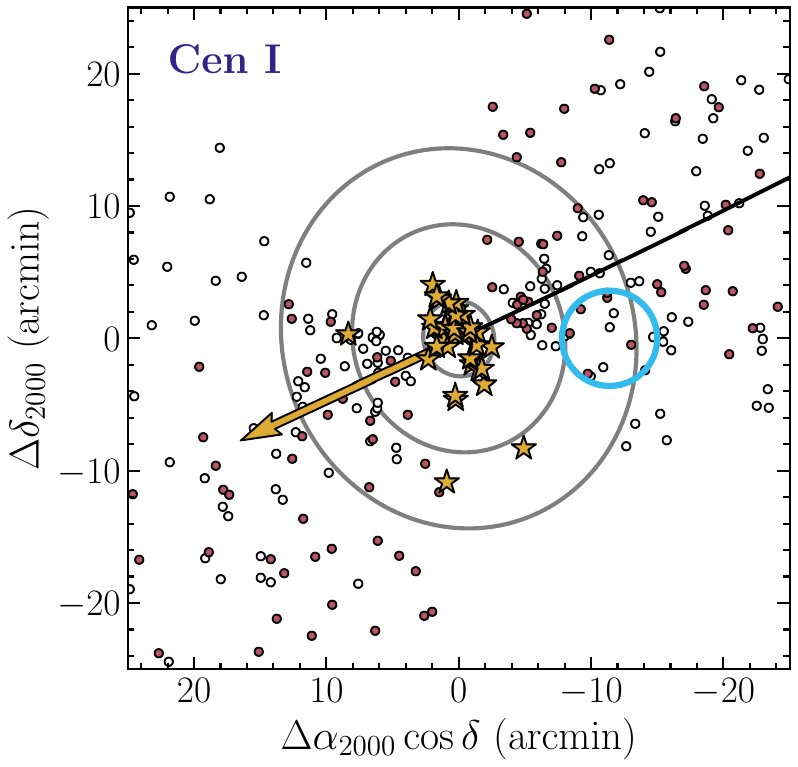}
\caption{Particle-spray models for Eri~IV (Left) and Cen~I (Right). The yellow stars mark spectroscopically confirmed members and the yellow arrow is the systemic proper motion. Possible overdensities previously identified from photometric data are circled in blue. The grey ellipses indicate 1, 3, and 5 half-light radii. The pink circles are the sampled stars from the particle-spray model including the effect of the LMC on the Milky Way barycentric acceleration, and the open circles are the sampled stars when this effect is not considered.
\label{fig:streams}}
\end{figure*}

We also investigated the possibility of tidal stripping for Eri~IV and Cen~I. The discovery papers for each pointed out possible overdensities that could be tidal features (see Figure 1 of \citealt{2021ApJ...920L..44C} and Figure 1 of \citealt{2020ApJ...890..136M}). Additionally, the high ellipticity like that of Eri~IV ($\varepsilon = 0.54^{+0.10}_{-0.14}$) has been suggested to evince tidal disruption (e.g., \citealt{2010ApJ...721.1333M, 2010AJ....140..138M}, although see \citealt{Goater2023:2023arXiv230705130G}). We modelled tidal streams of each system using \texttt{streamspraydf} \citep{2015MNRAS.452..301F}, as implemented in \texttt{galpy} \citep{2015ApJS..216...29B,2022MNRAS.511.2339Q}, which is a particle-spray model of the distribution function of a tidal stream for a given system. We use the enclosed masses measured in the previous section, set the start of the disruption at approximately the most recent pericenter passage, and sample 300 stars in each of the leading and trailing arms. The present-day locations of the model stream stars are presented in Figure \ref{fig:streams}. The approximate locations of the proposed tidal features are circled in blue on each figure.

As with the orbit, the stream is modelled with the default potential plus the effect of the LMC on the Milky Way center's barycentric acceleration and again with the default potential alone, excluding the effect of the LMC. The inclusion of the reflex motion of the Milky Way rotates the streams, an effect previously demonstrated for dwarf galaxy Antlia II \citep{2021ApJ...921...32J}. The effect of a non-static potential has also been widely discussed in literature \citep[e.g.,][]{2015ApJ...802..128G,Garavito-Camargo2021:2021ApJ...919..109G,2022MNRAS.512..739D}. The rotation here is somewhat more obvious with Cen~I, which has a less diffuse stream model than Eri~IV. The effect is very subtle on the scale of each of the proposed overdensities and becomes pronounced only on a scale of a few degrees, and so unlike Antlia II, the morphology and proper motion of Eri~IV and Cen~I are not uniquely explained by models that include the effect of the LMC on the barycentric acceleration of the Milky Way.

We do find that the overdensity proposed in the discovery paper of Eri~IV is a true feature, as five spectroscopic members lie in that region. The exact nature of this feature is unclear. It is possible that selection effects cause this feature to appear overdense, as we specifically targeted possible tidal features and did not target many objects opposite this specific feature. Additionally, while the stream model does not preclude tidal stripping as a cause of this feature, we do not see any additional evidence to suggest that Eri~IV is being tidally disrupted. 

Tidally disrupting systems are expected to have, for example, clear tidal tails, like those seen around Tucana III \citep{Drlica-Wagner2015:2015ApJ...813..109D}, a large velocity gradient, particularly in the direction of orbit \citep[e.g.,][]{2018ApJ...866...22L, 2021ApJ...921...32J}, and/or a tidal radius comparable to the half-light radius ($r_t/r_{1/2}\approx1$) \citep{2022ApJ...940..136P}. 

\begin{figure}
\includegraphics[width = \columnwidth]{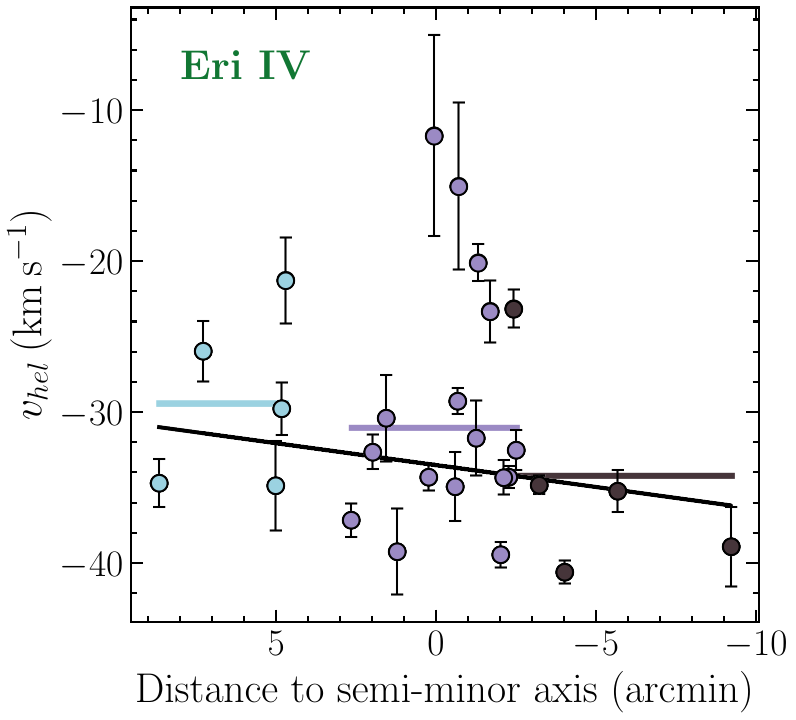}
\caption{Velocity gradient of Eri~IV along the semi-major axis. The member stars are divided into three groups: stars within one half-light radius (purple), stars outside of one half-light radius and above the semi-minor axis (blue), and stars outside of one half-light radius and below the semi-minor axis (brown). The systemic velocity of the subgroup is indicated by a solid line of the same color, and solid black line is the line of best fit to the entire sample. The stars in blue are the same as those in the overdensity circled in Figure \ref{fig:streams}. We note that the two apparent outliers at $v_{hel} \approx -15 \:\mathrm{km\:s^{-1}}$ with large error bars are the BHBs discussed in Section \ref{sec:eri4mems}.
\label{fig:gradient}}
\end{figure}

We measure the velocity gradient of Eri~IV using a multivariate Gaussian model and sample using \texttt{emcee} to simultaneously fit $v_{hel}$, $\sigma_{v}$, the position angle $\phi$, and the gradient $dv_{hel}/d\chi$, where $\chi$ is the angular distance of a star from the center of the system at position angle $\phi$ that maximizes the gradient (see \citealt{2017ApJ...838....8L} for more detail on the model). No signal is recovered for the position angle $\phi$ and the gradient is consistent with 0. We also divided the sample into three groups along the semi-major axis (stars outside of $1\:r_{h}$ and below the semi-minor axis, stars within $1\:r_{h}$, and stars outside of $1\:r_{h}$ and above the semi-minor axis) and calculated the systemic parameters of these groups (see Figure \ref{fig:gradient}). The difference in systemic velocity between stars on the left and right hand side of Eri~IV is $4.8\pm4.0\:\mathrm{km\:s^{-1}}$, which corresponds to a gradient of $0.5\pm0.5\:\mathrm{km\:s^{-1}\:arcmin^{-1}}$ along the semi-major axis. This also indicates that Eri~IV does not have a detectable velocity gradient. Additionally, if the tidal radius of Eri~IV is approximated as the Jacobi radius \citep{2008gady.book.....B} and  $M_{1/2}$ is as measured above, we estimate its tidal radius at pericenter is $r_t = 498\pm128\:\mathrm{pc}$, decidedly beyond $5\:r_{1/2}$ ($r_t/r_{1/2} \approx 6.6\pm0.3$) and already likely an underestimate. Overall, the lack of a strong velocity gradient and the large tidal radius indicate that Eri~IV is unlikely to be tidally disrupting. 

None of the observed member stars of Cen~I lie in the proposed overdensity (it was not covered by any of our masks), so we cannot confirm whether it is a true feature, but the stream model does predict stripped stars in that region. However, the presence of spray stars in these regions does not indicate tidal stripping---their absence might suggest that tidal stripping is unlikely, but their presence does not confirm it. There is also no evidence of a velocity gradient in Cen~I. Along the semi-major axis, the velocity gradient is $-0.3\pm 0.4 \:\mathrm{km\:s^{-1}\:arcmin^{-1}}$, which is consistent with 0 within 1$\sigma$. The tidal radius of Cen~I is estimated as approximately $r_t = 369 ^{+158}_{-106}\:\mathrm{pc}$ at its pericenter passage. This tidal radius is comparable to $5\:r_{1/2}$ ($r_t/r_{1/2}\approx4.7\pm0.5$), which might suggest that the stars on the very outskirts of Cen~I are vulnerable to tidal stripping. However, the tidal radius used here is likely an underestimate, and so, combined with the lack of a velocity gradient, we conclude that our observations show no evidence of tidal stripping of Cen~I.

Overall, further observations, particularly of the outskirts of both galaxies, would help clarify the nature of the features highlighted in Figure \ref{fig:streams} and improve our interpretation of their morphological and orbital parameters with respect to tidal disruption.

\subsection{Metallicity Distribution Function of Eri~IV}\label{sec:mdf}
The metallicity distribution function (MDF) of Eri~IV is notable in two ways. First, it has an unusually low mean, compared to other dwarf galaxies of similar stellar mass (Figure \ref{fig:mvcomparison}). Second, nearly all of the stars are concentrated in a low-metallicity peak with $\mbox{[Fe/H]} \sim -3.0$, with only one star at $\mbox{[Fe/H]} \sim -2.0$ (Figure \ref{fig:eri4hist}).

Though Eri~IV slightly deviates from the empirical luminosity-metallicity relation, it lines up with an extrapolation of the power law fit for brighter systems to lower luminosity, in contrast to the bulk of UFDs that form an apparent [Fe/H] plateau (Figure \ref{fig:mvcomparison} shows the absolute magnitude-metallicity relationship, which also displays the plateau). This single power law is produced by the vast majority of theoretical models, where it is a sensitive probe of the mass-dependence of stellar feedback in a UFD \citep[e.g.,][]{2017MNRAS.472.2356M,2019MNRAS.490.4447W,2020MNRAS.491.1656A,2022arXiv220913663W}.

The [Fe/H] plateau can be interpreted in three ways:
(1) the majority of these UFDs originally lay on the luminosity-metallicity relation but have since lost stellar mass through tidal disruption \citep[e.g.,][]{2013ApJ...770...16K}; 
(2) there is some additional physics creating the plateau, such as Pop~III pre-enrichment \citep{2019MNRAS.490.4447W} or a ceiling in stellar feedback \citep{2022MNRAS.514.2667K};
(3) there are significant observational biases in current UFD mean metallicities, potentially due to sample sizes \citep{Fu2023:2023arXiv230606260F}, unexplored radial metallicity gradients \citep[e.g.,][]{2021NatAs...5..392C}, and/or significantly underestimated uncertainties, that produce the appearance of a plateau.

It is likely that all of these explanations contribute in some way, but Eri~IV's consistency with the extrapolated power law suggests to us that the plateau is best explained by observational effects and/or additional physical processes.
With respect to observational effects, it is possible that those galaxies that appear to lie along the plateau are a result of small number statistics (or other biases). First, the mean metallicities of ultra-faint dwarfs are at times determined by only a handful of the brightest stars due to the difficulty inherent to observing such faint systems, but the uncertainties reported may not reflect the magnitude of possible sampling errors introduced by such small samples. There may also be other unaccounted-for uncertainties. Additionally, the sample of galaxies at such faint magnitudes is unlikely to be complete, although there is no expected bias towards more metal-rich systems in particular. 

Setting aside possible observational biases or issues, Eri~IV disfavors tidal disruption as the dominant source of this effect. Eri~IV is likely not tidally disrupting, and has orbital parameters (Figure \ref{fig:apoperiecc}) and $r_t/r_{1/2}$ comparable to the majority of the UFDs on the plateau \citep{2022ApJ...940..136P}. Thus, actual physical differences in metallicity floors (due to stochastic pre-enrichment by Pop~III stars \citep[e.g.,][]{2015MNRAS.454..659J}) or stellar feedback and star formation histories between different UFDs seems plausible.

Furthermore, it is interesting to consider the shape of the Eri~IV MDF (shown in Figure \ref{fig:eri4hist}). The vast majority of chemical evolution models produce left-skewed metallicity distributions \citep[e.g.,][]{2012ApJ...751...46K, 2022arXiv221017045S}, as metallicity enrichment is roughly linear and [Fe/H] is a log quantity. The MDF of Eri~IV, however, appears to be right-skewed: nearly all stars are concentrated in a low metallicity $\mbox{[Fe/H]}=-3$ peak, but there is a tail to high metallicity $\mbox{[Fe/H]}=-2$.

It is worth asking if the observed skew is real and physical or merely the result of a small sample size or observational bias. 
We see no evidence of a radial metallicity gradient nor a gradient along the semi-minor axis (both are consistent with zero), so our target selection is unlikely to have introduced a metallicity-related bias. 
The sample size for Eri~IV of 16 stars is comparable to that of other UFDs and relatively robust to potential contamination, but it is still small enough to question the significance of the skew. To this end, we conducted an Anderson-Darling test and a Monte Carlo simulation. 

The Anderson-Darling test is a non-parametric test that evaluates whether a sample was drawn from a reference distribution. While not a test of asymmetry specifically, it is a powerful test of normality, even when the sample size is small \citep{1974stephens}. For Eri~IV, the Anderson-Darling test shows that the hypothesis that the samples are drawn from a normal distribution can be rejected with 97.5\% confidence. For comparison, the same test on Cen~I shows no statistically significant departure from normality. 

We then compared the proportion of random 16-star samples that have comparable skewness to the observed sample when drawn from a normal distribution versus a skew-normal distribution. We quantify skewness using the adjusted Fisher-Pearson moment coefficient of skewness. For additional detail on this test, see Appendix \ref{sec:appendix}. We find that only 0.5\% of samples drawn from a normal distribution can reproduce the observed skew, compared to 8.2\% of the samples drawn from a skew-normal. 

These results and the magnitude of the observed skew suggest that it may be physical rather than an artifact of sample size. We also note that we have compared the observed MDF to a normal distribution. If we were to assume that the MDF is left-skewed, as predicted by most models, the observed right skew would be even less likely to arise by chance.

Although atypical, right-skewed MDFs have been seen before, most notably in the dwarf galaxy Ursa Minor \citep{2011ApJ...727...78K}. Indeed, the right-skewed MDF in Ursa Minor is likely a result of peculiar chemical enrichment that is dominated by Type~Ia supernovae. \citet{2018ApJ...857...97M} used data from \citet{2010ApJ...719..931C} to show that the most Fe-rich stars in Ursa Minor are highly enriched by Type~Ia supernovae. One possible scenario is that the most Fe-rich stars in Ursa Minor formed out of gas with extreme inhomogeneous metal mixing, potentially forming directly out of gas swept up by the Type Ia supernova remnant \citep[e.g.,][]{2019MNRAS.482.1304E}.
We suggest this same mechanism could have occurred in Eri~IV, and it should be easily tested with detailed chemical abundances. 

\subsection{Astrophysical J-factor and D-factor}\label{sec:jfactor}

\begin{figure*}\centering
\includegraphics[width =0.9\textwidth]{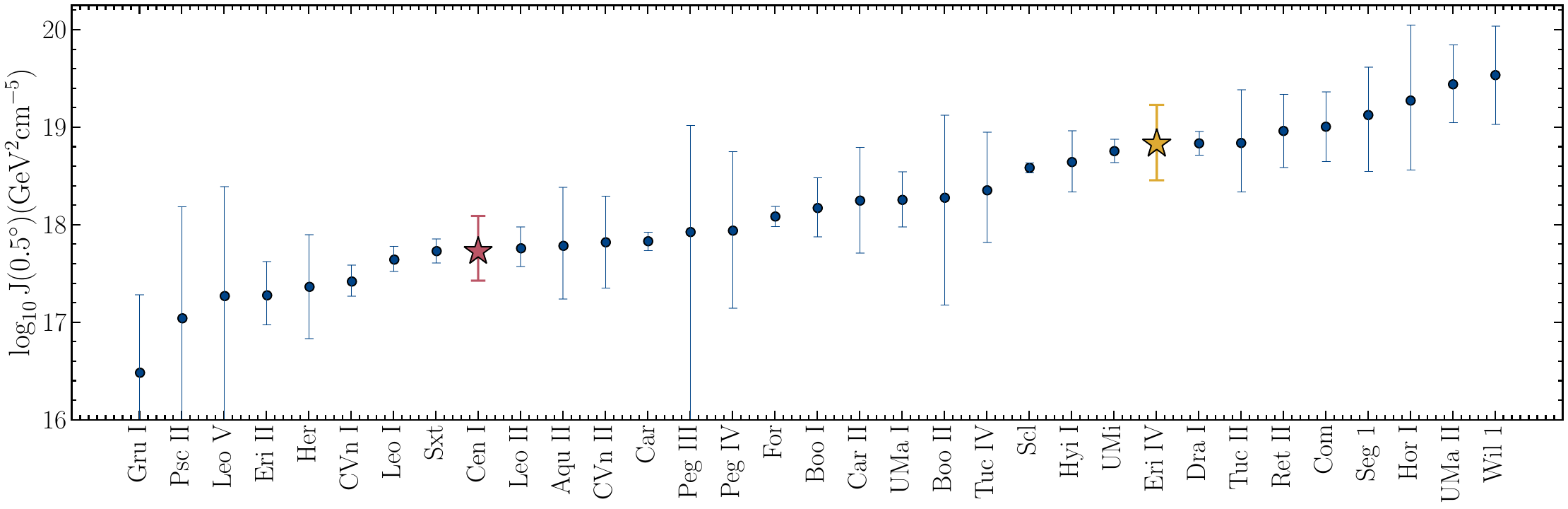}
\caption{J-factor comparison of Eri~IV (yellow star), Cen~I (pink star), and other Milky Way dwarf galaxies with a resolved velocity dispersion (blue circles) from \protect \citet{Pace2019MNRAS.482.3480P, Boddy2020PhRvD.102b3029B, Chiti2022ApJ...939...41C, 2023ApJ...942..111C, 2023arXiv230203708B}. Some large J-factors in the literature should be treated with caution; see text for details.
\label{fig:jfac}}
\end{figure*}

The Milky Way satellite dwarf spheroidal galaxies are excellent targets to search for dark matter annihilation or decays products due to their close proximity, high dark matter fractions, and low intrinsic backgrounds \citep{2015PhRvL.115w1301A}. Any dark matter annihilation or decay signal depends on the particle physics properties (e.g., annihilation cross section and particle mass) and the properties of the astrophysical environment (the dark matter density distribution). Specifically, this second astrophysical component of the annihilation signal, called the J-factor, is the line-of-sight integral through the dark matter density distribution integrated over a solid angle. The equivalent for the decay signal is referred to as the D-factor. In conjunction with gamma-ray observations, the J-factor can be used to constrain the microphysical properties of dark matter \citep{2015PhRvL.115w1301A}. A higher J-factor implies that a galaxy would have a stronger annihilation signal, and would therefore be a better site for effective indirect detection.

The standard methodology to compute the J-factor for the dSphs is to compare solutions from the spherical Jeans equation to the observed velocity dispersion to infer the underlying properties of the dSph dark matter halo \citep[e.g..][]{GeringerSameth2015ApJ...801...74G, Bonnivard2015MNRAS.453..849B, Pace2019MNRAS.482.3480P}. 
We follow \citet{Pace2019MNRAS.482.3480P} to compute the J-factor, which assumes the stellar distribution to follow a Plummer model \citep{Plummer1911MNRAS..71..460P}, an NFW model for the dark matter halo \citep{Navarro1996ApJ...462..563N}, and constant stellar anisotropy with radius. 

We apply this analysis to the 26 and 32 non-variable star samples for Eri~IV and Cen~I, respectively. 
For the J-factor of Eri~IV, we find $\log_{10}{J(\theta)}=18.5\pm0.4$, $18.6^{+0.4}_{-0.3}$,  $18.8\pm0.4$, and $18.9\pm0.4$   for solid angles of $\theta=0.1^{\circ}, 0.2^{\circ}, 0.5^{\circ}, 1^{\circ}$ in logarithmic units of ${\rm GeV^{2}~cm^{-5}}$.
For  Cen~I, we find $\log_{10}{J(\theta)}=17.6\pm0.3$, $17.7\pm0.3$,  $17.7_{-0.3}^{+0.4}$, and $17.8_{-0.3}^{+0.4}$   for solid angles of $\theta=0.1^{\circ}, 0.2^{\circ}, 0.5^{\circ}, 1^{\circ}$ in logarithmic units of ${\rm GeV^{2}~cm^{-5}}$. 
For the D-factor of Eri~IV, we find $\log_{10}{D(\theta)}=17.4\pm0.2$, $17.8\pm0.3$, and $18.3_{-0.5}^{0.4}$,   for solid angles of $\theta=0.1^{\circ}, 0.2^{\circ}, 0.5^{\circ}$ in logarithmic units of ${\rm GeV~cm^{-2}}$.
For  Cen~I, we find $\log_{10}{D(\theta)}=16.8_{-0.2}^{+0.3}$, $17.2_{-0.3}^{+0.4}$, and $17.5_{-0.5}^{+0.6}$ for solid angles of $\theta=0.1^{\circ}, 0.2^{\circ}, 0.5^{\circ}$ in logarithmic units of ${\rm GeV~cm^{-2}}$. 

In Figure~\ref{fig:jfac}, we compare the J-factors of Eri~IV and Cen~I to the dSph population. The literature comparison sample uses the same J-factor methodology and the values are taken from \citet{Pace2019MNRAS.482.3480P, Boddy2020PhRvD.102b3029B, Chiti2022ApJ...939...41C, 2023ApJ...942..111C, 2023arXiv230203708B}.
We note that some large J-factors in the literature should be treated with caution; for example, Wil~1 may be tidally disrupting \citep{2011AJ....142..128W} and has a small pericenter \citep[$\sim12 ~{\rm kpc}$][]{2022ApJ...940..136P}, the velocity dispersion of Tuc~II has decreased by a factor of two with larger samples \citep{2023AJ....165...55C}, different stellar density profiles of Hor~I lead to a factor of two uncertainty in the J-factor \citep{Pace2019MNRAS.482.3480P}, and  membership choices of the spectroscopic sample impact the Seg~1 J-factor \citep{2015MNRAS.453..849B}.   
Compared to the general dSph population, Eri~IV  has a relatively large J-factor whereas Cen~I is unremarkable as it is similar to dSphs at similar distances. 
The predicted J-factor of Eri~IV and Cen~I is $\log_{10}{J(0.5^{\circ})}\sim 18.6$ and $\sim 17.5$ based on the velocity dispersion, distance, and half-light radius \citep{Pace2019MNRAS.482.3480P}. The predictions of both galaxies are $\sim 0.2~{\rm dex}$ smaller than the full dynamical analysis. The scaling also examines the overall difference in J-factor between the two galaxies. A factor of $\sim10.8\sim(6.14/4.17)^4/(76.7/116.3)^2$ difference in J-factor is predicted based on the difference in velocity dispersion and distance between the two galaxies. 
Eri~IV is an excellent target for dark matter indirect detection searches while Cen~I will be another useful target for stacked analysis.

\section{Summary}\label{sec:summary}
Using medium resolution IMACS/Magellan spectroscopy, we identified 28 member stars of Eri~IV and 34 member stars of Cen I, two Milky Way satellites first identified in the DELVE survey. Using the infrared Calcium triplet lines, we measured the velocities and metallicities of member stars. We also resolve velocity and metallicity dispersions for both systems. Based on their physical size and mass-to-light ratio, we confirm the classification of both systems as dwarf galaxies.

Specifically, we find that Eri~IV has a velocity dispersion of $\sigma_{v} = 6.1^{+1.2}_{-0.9}\:\mathrm{km\:s^{-1}}$ and a corresponding mass-to-light ratio of $M_{1/2}/L_{1/2} = 598^{+361}_{-228} M_{\odot}/L_{\odot}$. Using 16 Eri~IV members, we find that it has a metallicity of $\mathrm{[Fe/H]} = -2.87^{+0.08}_{-0.07}$ and dispersion of $\sigma_{\mathrm{[Fe/H]}} = 0.20\pm0.09$. When compared to other dwarf galaxies and globular clusters, the morphological and orbital parameters of Eri~IV are consistent with those of the dwarf galaxies. Our analysis also suggests that it is unlikely that Eri~IV is being tidally stripped. We have shown that these conclusions are unlikely to be influenced by the presence of an unidentified binary, variable star, or field star, with the possible exception of the most metal-rich member star. Should this star be misidentified and in fact be a field star, the metallicity of Eri~IV would drop to $-2.95$ dex and the dispersion to $0.04^{+0.07}_{-0.03}$. Importantly, this star has no effect on the conclusion that Eri~IV is in fact a dwarf galaxy, and that it is one of the most metal-poor dwarf galaxies ever discovered. This especially low metallicity, Eri~IV's somewhat unusual metallicity distribution, and the presence of many extremely metal-poor stars make it an appealing candidate for further high resolution spectroscopic follow-up. Eri~IV also has a large J-factor ($\log_{10}{\left(J(0.5^{\circ})/{\rm GeV^{2}~cm^{-5}} \right)}=18.8$), indicating that it would be a good target for indirect dark matter detection experiments.

Cen~I has a velocity dispersion of $\sigma_{v} = 4.2^{+0.6}_{-0.5}\:\mathrm{km\:s^{-1}}$ and mass-to-light ratio of $M_{1/2}/L_{1/2} = 140^{+60}_{-42} M_{\odot}/L_{\odot}$. We measure the systemic metallicity using 27 member stars and find $\mathrm{[Fe/H]} = -2.57\pm0.08$ and $\sigma_{\mathrm{[Fe/H]}} = 0.38^{+0.07}_{-0.05}$. Its morphological and orbital parameters are consistent with other dwarf galaxies, and we see no evidence of tidal stripping.

The classification of Eri~IV and Cen~I adds two dwarf galaxies to the rapidly growing sample of observed Milky Way satellite galaxies. Recent discoveries in surveys like DELVE and forecasts for future missions like the Vera Rubin Observatory indicate that the sample is far from complete \citep{2014ApJ...795L..13H,2019ApJ...873..111I, 2021ApJ...918...88M, 2022MNRAS.516.3944M}, and spectroscopy will be crucial to understanding and characterizing new faint stellar systems like Eri~IV and Cen~I. An improved, well-understood sample of Milky Way satellites and their member stars will invariably provide important constraints on the nature of dark matter, the evolution of the Milky Way, and early galaxy evolution.

\section{Acknowledgements}\label{sec:6}

M.E.H. \& T.S.L. acknowledge financial support from Natural Sciences and Engineering Research Council of Canada (NSERC) through grant RGPIN-2022-04794.
A.B.P. is supported by NSF grant AST-1813881. 
J.D.S. is supported by NSF grant AST-1714873.

This paper includes data gathered with the 6.5 meter Magellan Telescopes located at Las Campanas Observatory, Chile.

This project used data obtained with the Dark Energy Camera (DECam), 
which was constructed by the Dark Energy Survey (DES) collaboration.
Funding for the DES Projects has been provided by 
the DOE and NSF (USA),   
MISE (Spain),   
STFC (UK), 
HEFCE (UK), 
NCSA (UIUC), 
KICP (U. Chicago), 
CCAPP (Ohio State), 
MIFPA (Texas A\&M University),  
CNPQ, 
FAPERJ, 
FINEP (Brazil), 
MINECO (Spain), 
DFG (Germany), 
and the collaborating institutions in the Dark Energy Survey, which are
Argonne Lab, 
UC Santa Cruz, 
University of Cambridge, 
CIEMAT-Madrid, 
University of Chicago, 
University College London, 
DES-Brazil Consortium, 
University of Edinburgh, 
ETH Z{\"u}rich, 
Fermilab, 
University of Illinois, 
ICE (IEEC-CSIC), 
IFAE Barcelona, 
Lawrence Berkeley Lab, 
LMU M{\"u}nchen, and the associated Excellence Cluster Universe, 
University of Michigan, 
NSF's National Optical-Infrared Astronomy Research Laboratory, 
University of Nottingham, 
Ohio State University, 
OzDES Membership Consortium
University of Pennsylvania, 
University of Portsmouth, 
SLAC National Lab, 
Stanford University, 
University of Sussex, 
and Texas A\&M University.

The DELVE project is partially supported by the NASA Fermi Guest Investigator Program Cycle 9 No. 91201. This work is partially supported by Fermilab LDRD project L2019-011.

This work has made use of data from the European Space Agency (ESA) mission
\textit{Gaia} (\url{https://www.cosmos.esa.int/gaia}), processed by the \textit{Gaia}
Data Processing and Analysis Consortium (DPAC,
\url{https://www.cosmos.esa.int/web/gaia/dpac/consortium}). Funding for the DPAC
has been provided by national institutions, in particular the institutions
participating in the \textit{Gaia} Multilateral Agreement.

C.E.M.-V. is supported by the international Gemini Observatory, a program of NSF's NOIRLab, which is managed by the Association of Universities for Research in Astronomy (AURA) under a cooperative agreement with the National Science Foundation, on behalf of the Gemini partnership of Argentina, Brazil, Canada, Chile, the Republic of Korea, and the United States of America.

W.C. gratefully acknowledges support from a Gruber Science Fellowship at Yale University.

J.A.C.-B. acknowledges support from FONDECYT Regular N 1220083.

\appendix
\section{Statistical tests of skewness}\label{sec:appendix}

We provide here additional detail on the statistical tests conducted in Section \ref{sec:mdf}, the Anderson-Darling test and a Monte Carlo simulation.

The Anderson-Darling test was chosen over other similar statistical tests like a Kolmogorov-Smirnov test because it gives more weight to the tails of the distributions and has greater statistical power at small sample sizes while remaining flexible \citep{Razali2011powercomparisons}. The data are standardized and compared to a reference distribution, $\mathcal{N}(0,1)$. A Shapiro-Wilk test, which only tests for normality but is the most powerful such test \citep{Razali2011powercomparisons}, was also performed. The Shapiro-Wilk test indicates that hypothesis that the data are drawn from a normal distribution can be rejected at 97.5\% confidence, in agreement with the results of the Anderson-Darling test \citep{Shapiro_Wilk_1965}.

With respect to the Monte Carlo simulation, skewness is quantified using the adjusted Fisher-Pearson moment coefficient of skewness, $G_1$. There is no set value of $|G_1|$ above which a distribution is considered skewed, so we assume a conventional value of $|G_1| > 1$ \citep{Bulmer1967.book.....B}. For Eri~IV, $G_1 = 1.49$, indicating that it is right-skewed. For comparison, $G_1 = 0.09$ for Cen~I. 

The skew-normal distribution for all draws is described by the best-fit parameters for the observed sample, estimated with the maximum-likelihood estimator implemented in \texttt{scipy.rv\_continuous}. The normal distribution for each draw is described by a mean metallicity and dispersion sampled from their error distributions. We drew 10,000 16-star samples from each of these distributions. Approximately 2.6\% (23.1\%) of the samples drawn from normal (skew-normal) distributions are right skewed ($G_1 > 1$) and 0.5\% (8.2\%) have have a skew as extreme as the observed sample ($G_1 > 1.4$). 

These proportions are unsurprising, as the main driver of the observed skewness is the metal-rich member star, the metallicity of which is approximately $2.8\sigma$ from the mean, where only a few percent of stars in Eri~IV would lie if the MDF were normal. When we exclude the most metal-rich member star, the coefficient of skewness of the observed sample is $G_1 = 0.84$, which could indicate a mild right-skew, but we do not confidently make that determination. Approximately 5\% (33\%) of the 10,000 samples from the normal (skew-normal) distribution have skewness at least as extreme as $G_1 = 0.84$, and the Anderson-Darling test indicates that this 15-star sample has no meaningful departure from normality. 

The sampling procedure also assumes the observed stars are sufficiently representative of the population. This has not always held for other UFDs---mean metallicities of several UFDs have changed considerably upon further analysis with larger spectroscopic samples \citep[e.g.,][]{Chiti2022ApJ...939...41C, 2023arXiv230203708B}. Our sample is 2-3 times larger than most such initial spectroscopic samples, so a non-representative sample is less likely, although certainly possible.

\bibliography{heiger_imacs_spectroscopy_of_eri4_and_cen1}{}
\bibliographystyle{aasjournal}

\end{document}